\documentclass[12pt]{iopart}

\usepackage{graphicx}
\usepackage{color}
\usepackage{soul}
\usepackage{url}
\usepackage{bm}         
\usepackage{times}
\usepackage{dcolumn}
\usepackage{bm}
\usepackage{epsf}
\usepackage{amssymb}
\usepackage{booktabs}
\usepackage{tensor}
\usepackage{siunitx}
\usepackage[caption=false]{subfig}
\usepackage[%
  bookmarksopen=true,
  colorlinks=true,
  urlcolor=blue,
  linkcolor=blue,
  citecolor=blue
]{hyperref}

\usepackage[normalem]{ulem}

\begin{document}
\title[Late-time post-merger modeling of a compact binary]{Late-time post-merger modeling of a compact binary:  effects of relativity, r-process heating, and treatment of transport effects}

\author{
Milad Haddadi$^{1}$
Matthew D. Duez$^{1}$,
Francois Foucart$^{2}$,
Teresita Ramirez$^{2,7,8}$,
Rodrigo Fern\'andez$^{3}$,
Alexander L. Knight$^2$,
Jerred Jesse$^{1}$,
Francois H\'{e}bert$^{4}$,
Lawrence E. Kidder$^5$,
Harald P. Pfeiffer$^6$,
Mark A. Scheel$^4$
}

\address{$^1$ Department of Physics \& Astronomy, Washington State University, Pullman, Washington 99164, USA}
\address{$^2$ Department of Physics \& Astronomy, University of New Hampshire, 9 Library Way, Durham NH 03824, USA}
\address{$^3$ Department of Physics, University of Alberta, Edmonton, AB T6G 2E1, Canada}
\address{$^4$ TAPIR, Walter Burke Institute for Theoretical Physics, MC 350-17, California Institute of Technology, Pasadena, California 91125, USA}
\address{$^5$ Cornell Center for Astrophysics and Planetary Science, Cornell University, Ithaca, New York, 14853, USA}
\address{$^6$ Max Planck Institute for Gravitational Physics (Albert Einstein Institute), Am M{\"u}hlenberg~1, D-14476 Potsdam, Germany}
\address{$^7$ Gravitational Wave Physics and Astronomy Center, California State University Fullerton, Fullerton, California 92834, USA}
\address{$^8$ Center for Interdisciplinary Exploration and Research in Astrophysics (CIERA), Physics and Astronomy, Northwestern University, Evanston, Illinois 60202, USA}
\begin{abstract}
Detectable electromagnetic counterparts to gravitational waves from compact binary mergers can be produced by outflows from the black hole-accretion disk remnant during the first ten seconds after the merger.  Two-dimensional axisymmetric simulations with effective viscosity remain an efficient and informative way to model this late-time post-merger evolution.  In addition to the inherent approximations of axisymmetry and modeling turbulent angular momentum transport by a viscosity, previous simulations often make other simplifications related to the treatment of the equation of state and turbulent transport effects.

In this paper, we test the effect of these modeling choices.  By evolving with the same viscosity the exact post-merger initial configuration previously evolved in Newtonian viscous hydrodynamics, we find that the Newtonian treatment provides a good estimate of the disk ejecta mass but underestimates the outflow velocity.  We find that the inclusion of heavy nuclei causes a notable increase in ejecta mass.  An approximate inclusion of r-process effects has a comparatively smaller effect, except for its designed effect on the composition.  Diffusion of composition and entropy, modeling turbulent transport effects, has the overall effect of reducing ejecta mass and giving it a speed with lower average and more tightly-peaked distribution.  Also, we find significant acceleration of outflow even at distances beyond 10,000\,km, so that thermal wind velocities only asymptote beyond this radius and at higher values than often reported.

\end{abstract}

\maketitle

\section{Introduction} \label{sec:introduction}

Both binary neutron star mergers and black hole-neutron star mergers have been detected in gravitational waves~\cite{LIGOScientific:2017vwq,LIGOScientific:2020aai,LIGOScientific:2021qlt}, and in one case counterpart signals were seen throughout the electromagnetic spectrum.  Part of the electromagnetic signal from these mergers will be associated with the dynamical ejecta released during the merger, but part is caused by jets and disk winds produced in the first seconds after the merger.  The properties of a kilonova, in particular, depend on the mass of ejected material and its average velocity and composition, usually parametrized in simulations by the fraction of the baryonic mass in protons, $Y_e$~\cite{Lippuner2015,Metzger:2019zeh}.

Modeling the multi-second evolution of post-merger systems presents several numerical challenges.  The orbital period of the inner accretion disk will be on the order of a millisecond, so one must evolve around $10^4$ dynamical times.  A similar number of decades of spatial scales must be modeled, ranging from the black hole horizon radius to the outgoing ejecta.  Crucial physical processes are difficult to model, and may be treated with varying levels of adequacy.  Angular momentum transport in the disk material must be included either through direct modeling of magnetorotational turbulence or by an effective viscosity prescription.  Neutrinos importantly effect the thermal and composition evolution for the first few hundred milliseconds, so neutrino radiation transport must be included.  In simple leakage models, neutrino radiation fields are not evolved, but estimated local cooling rates of the gas are included.  Moment closure schemes like M1 evolve angular moments of the neutrino distribution functions, truncated with a stipulated closure condition.  Most ambitious and adequate is to evolve the full distribution functions on 6D phase space.  Finally, simulations must be able to handle a wide range of densities ($1$--$10^{12}$ g cm${}^{-3}$) and temperatures ($10^7$--$10^{10}$K) and so, in principle, a wide range of nuclear reaction rates.  At high densities and temperatures, strong nuclear reactions are fast compared to dynamical times and settle to equilibrium, so isotope abundances and the equation of state are set (for a given proton-to-neutron ratio) by nuclear statistical equilibrium (NSE).  At lower densities and temperatures, this can no longer be assumed.

Only three-dimensional relativistic magnetohydrodynamic (MHD) simulations can treat the angular momentum transport entirely {\it ab initio}.  Impressively, a small number of such simulations of neutrino-cooled accretion disks around black holes have been carried out~\cite{2018ApJ...858...52S,2017PhRvL.119w1102S,10.1093/mnras/sty2932,PhysRevD.100.023008,2021arXiv211104621H}, including a few which proceed for multi-second timescales~\cite{10.1093/mnras/sty2932,2021arXiv211104621H}, long enough to observe both an early-time, neutrino-rich fast outflow and a late-time ``thermal'' neutrino-poorer and slower outflow.

In order to explore efficiently the parameter space of possible post-merger states, it remains useful to carry out 2D axisymmetric simulations.  Because of the impossibility of an axisymmetric dynamo~\cite{1933MNRAS..94...39C}, late-time 2D MHD simulations must include explicit dynamo terms to incorporate non-axisymmetric or subgrid effects, and first simulations of this sort have in fact been carried out~\cite{Shibata:2021xmo}.  Most 2D simulations, though, have incorporated angular momentum transport through a Shakura-Sunyaev ``alpha'' viscosity~\cite{shakura:1973}.  This includes the Newtonian simulations of Fernandez and collaborators~\cite{Fernandez2013,Fernandez:2014,Fernandez:2014b,Fernandez:2016sbf,Fernandez:2020oow}, which identified the importance and rough properties of late-time outflows.  These simulations included a pseudo-Newtonian potential (to incorporate some effects of general relativity), a Helmholtz equation of state with free nucleons and alpha particles in assumed nuclear statistical equilibrium, and neutrino effects modeled by leakage emission and lightbulb-type self-irradiation.  Improvements to the treatment of some aspects of the physics were made by Just~{\it et al}~\cite{10.1093/mnras/stv009} and Fujibayashi~{\it et al}~\cite{PhysRevD.101.083029,Fujibayashi:2020jfr}.  Both included genuine neutrino transport via moment closure schemes and general relativity (the former in the conformal flatness approximation, the latter in full general relativity).  Newtonian and relativistic simulations have shown largely consistent results for disk and outflow properties.  They are also consistent with MHD simulations, with the exception that only MHD simulations show the early, fast outflow component and, of course, Blandford-Znajek type energy outflows.

Most, but not all, 2D simulations use artificial initial data (e.g. tori with constant entropy per baryon, $Y_e$, and angular momentum).  This is sensible as a strategy for covering parameter space, and given the artificiality of azimuthally averaging merger simulation data, but the early (first few hundred milliseconds) evolution will presumably be sensitive to the initial state.  In one study, Fernandez~{\it et al}~\cite{Fernandez:2016sbf} used data from a 3D numerical relativity simulation of a black hole-neutron star merger as initial data.  (A previous set of 2D simulations had used a Newtonian 3D merger simulation for initial data~\cite{Fernandez:2014bra}.)  This is then a particularly useful case for comparing Newtonian to relativistic treatments and leakage to moment closure neutrino models, and we use it in this paper.  A few other instances of mapping merger simulation outcomes to post-merger simulations have been carried out~\cite{Fujibayashi:2017puw,Armengol:2021mbt}.

In fact, there are other modeling assumptions besides those related to gravity and neutrino transport that are worth checking.  First, there is the treatment of the equation of state.  Within the NSE regime, is it adequate to include only free nucleons and alpha particles?  Also, it is known that NSE will quickly become inaccurate far from the black hole, and that heating from r-process nucleosynthesis in particular may significantly affect outflows and fallback on these timescales~\cite{10.1111/j.1365-2966.2009.16107.x}.  Finally, because the physical mechanism of angular momentum transport modeled by viscosity is presumed to be turbulence, one would expect that same turbulence to generate other transport effects, such as composition and heat diffusion.  It has been suggested that if momentum transport is included in simulations, these others should be too, and in fact that they may affect outflows in interesting ways~\cite{Duez:2020lgq}.  We note that this is an uncertainty in the transport treatment independent of the well-known question of what value to set $\alpha$.

In this paper, we investigate the effect of these other modeling choices.  We carry out 2D relativistic viscous hydrodynamic simulations using moment closure neutrino transport using the Spectral Einstein Code (SpEC), proceeding from the same initial state used in Fernandez~{\it et al}~\cite{Fernandez:2016sbf}, namely the outcome of the merger of a 1.4\,$M_{\odot}$ neutron star with a 7\,$M_{\odot}$ black hole, simulated using the SpEC code.  Using this case allows us to perform a Newtonian-leakage vs relativistic-M1 comparison with no extra variables.  (Newtonian viscous hydrodynamics vs. relativistic MHD comparisons from an equivalent initial state have been performed previously~\cite{10.1093/mnras/sty2932}, but it is likely that MHD was the dominant cause of differences.)  Separate simulations include effects of heavy nuclei, r-process heating, and particle and entropy diffusion.

We find that for this case Newtonian simulations provide a good estimate of the total ejecta mass, but they underestimate the outflow velocity.  Treatment of the nuclear equation of state does have a significant effect on the outflow mass, in fact a much greater effect than r-process heating.  Composition and entropy transport effects tend to reduce the mass and velocity of ejecta.  Finally, we notice that asymptotic velocity must be calculated at farther distances from the source than is often done.

This paper is organized as follows.  In Section~\ref{sec:methods}, we present out evolution methods, noting particularly all alterations to the SpEC code to deal with long times, low densities, and low temperatures.  In Section~\ref{sec:results}, we present results and compare to the literature.  In Section~\ref{sec:discussion}, we present conclusions and discuss future work.

\section{Methods}
\label{sec:methods}

\subsection{Post-merger grid and initial data}

The initial state for simulations is the final state of a 3D black hole-neutron star simulation carried out in Foucart~{\it et al}~\cite{Foucart:2014nda}.  The simulation consists of a black hole with mass 7 $M_{\odot}$ and dimensionless spin $\chi=0.8$ (aligned with the orbital angular momentum of the binary) merging with an irrotational neutron star with mass 1.4 $M_{\odot}$.  During merger, the neutron star matter is modeled using the Lattimer-Swesty equation of state~\cite{Lattimer:1991nc} with nuclear incompressibility $K_0 = 220\,\mathrm{MeV}$, but by the end of the simulation, when only an accretion disk of density $\rho_0\lesssim 10^{12}$g cm${}^{-3}$ remains outside the black hole, nuclear forces outside nuclei have become unimportant (except for providing binding energy for nuclei).  This is one of the older black hole-neutron star simulations carried out using the Spectral Einstein Code, but we have deliberately chosen the same system as that evolved with a Newtonian code by Fernandez~{\it et al}~\cite{Fernandez:2016sbf}.  The post-merger simulation begins about 14\,ms after merger, at which time the disk has baryonic mass of about $M_0=0.07M_{\odot}$.

We evolve on a 2D spherical-polar grid.  In the asymptotically Minkowski coordinates inherited from the merger simulation, the grid extends in radius from $r_{\rm min}=26$\,km (close to the horizon) to $r_{\rm max}=120,000$\,km.  This outer radius, unusual for studies of this kind, allows us to study the evolution of ejecta velocity over a wide spatial range. The radial grid has 260 points.  To cover so many decades in radius, we use as our radial coordinate (in which the grid spacing is uniform) $u=\log(r)$.  The colatitude angular variable covers $0<\theta<\pi$ with 168 grid points.  We concentrate grid points near the equator by making the standard coordinate transformation
\begin{equation}
    \theta = \pi \zeta + \frac{1}{2}(1-h)\sin(2\pi \zeta)
\end{equation}
and evolving on a uniform grid in $\zeta$.  We choose $h=0.4$.  

Our resulting grid is comparable to that of Fernandez~{\it et al}~\cite{Fernandez:2016sbf}, with somewhat higher equatorial angular resolution but a bit lower radial resolution (in order to have a farther outer boundary with about the same number of radial points).  This grid size is fairly modest (smaller in radial resolution than Fernandez~{\it et al}'s more recent studies)~\cite{Fernandez:2020oow} and may well have errors of order 10\% in some quantities, but since the grid is constant across runs, we can readily identify the magnitudes and directions of effects of different physical inputs.

Converting from the 3D merger grid to the 2D post-merger grid requires azimuthally averaging.  The X-weighed azimuthal average of 3D function $f$ is
\begin{equation}
    \overline{f}(r,\theta) = \frac{\int_0^{2\pi}d\phi X(r,\theta,\phi)f(r,\theta,\phi)}{\int_0^{2\pi}d\phi X(r,\theta,\phi)}
\end{equation}
We average all metric functions with $X=1$:  the lapse $\alpha$, shift $\beta^i$, 3-metric $\gamma_{ij}$, and extrinsic curvature $K_{ij}$.  (Naturally, polar rather than Cartesian components must be averaged.)  From $\gamma_{ij}$ we know the determinant $\gamma=\det(\gamma_{ij})$.  We also average the density variable $\rho_{\star}$ with $X=1$, where $\rho_{\star}=\alpha\sqrt{\gamma}u^t\rho_0$, and $\rho_0$ is the rest (baryonic) density.  We compute averages of the electron fraction $Y_e$, temperature $T$, and covariant spatial components of the 4-velocity $u_i$ with weight $X=\rho$.  We have also tried weighting these averages by $\rho_0$ and averaging the pressure $P$ rather than the temperature [and then recovering $T(r,\theta)$ from the averaged $P$, $\rho_0$, and $Y_e$]; these sorts of choices have very little effect on the disk's initial global quantities and early evolution.

We do not add a fallback matter component, as did some of the runs in~\cite{Fernandez:2016sbf}.  The fallback is far from the black hole, making a Newtonian vs. relativistic comparison of its effects perhaps less interesting.  Also, the fallback component of these early runs was not tracked far and is not accurately known.  In future work, we will use more recent simulations where the bounded matter never leaves the merger grid.  (Fallback interaction with the disk makes the 10-20\,ms post-merger state of these newer runs less axisymmetric, so that azimuthal averaging might cause large initial disk oscillations.  One could avoid this by computing the azimuthal Reynolds stress$\sim \overline{\rho v_iv_j}-\overline{\rho}\overline{v}_i\overline{v}_j$ and adding this to the Euler equation.  However, one would need a model for how the azimuthal Reynolds stress decreases with time as the system axisymmetrizes.)

The post-merger grid must extend much farther than the merger grid.  The matter fields can be assumed to be vacuum outside of the original merger grid, but the metric must be extrapolated.  In particular, we extrapolate $\alpha-1$, $\gamma_{ij}-\delta_{ij}$, $\beta^i$, $K_{ij}$, which all fall off to zero as $r\rightarrow \infty$.  We find that simply setting $\alpha=M/r$ (even for a best fit of $M$) and setting the others to zero is insufficiently smooth and leads to observable artifacts when outflow crosses the merger grid outer boundary.  Instead, for each $\theta$, we use the last two radial points on the azimuthally averaged grid to fit each of the above metric coefficients to functions of the form $f_1(r) = A_1/r$ and $f_2(r) = A_2/r + B_2/r^2$.  We then use the one that falls off most quickly in $r$.  The resulting extrapolations appear at least visually smooth and reasonable, and we do not notice any artificial behavior when matter crosses into the extrapolated region.

To avoid unphysical disturbances from suddenly turning on viscous stresses, we activate the viscosity gradually and continuously, starting 7.5\,ms after the simulation begins and reaching its full strength about a millisecond later.

\subsection{Evolution methods}

\subsubsection{Hydrodynamics and handling of low densities}

We evolve the relativistic viscous hydrodynamic equations using the Spectral Einstein Code (SpEC)~\cite{SpEC2020} with multipatch methods for 2D axisymmetry as described in~\cite{Jesse:2020}.   (See~cite{Jesse:2020,Duez:2020lgq} for tests of our 2D viscous hydrodynamics and neutrino transport codes.)  The disk mass is less than a percent of the black hole mass even at the initial time, and the metric is very close to stationary in the merger coordinates, so we keep the spacetime metric fixed throughout the simulations.

The fluid is described by its baryonic rest density $\rho_0$, temperature $T$, electron fraction (really the proton fraction) $Y_e$, and 4-velocity $u^{\alpha}$.  The Lorentz factor is $W=\alpha u^t$.  The fluid has pressure $P$, specific internal energy $\epsilon$ and specific enthalpy $h=1+\epsilon+P/\rho_0$, which are related to $\rho_0$, $T$, and $Y_e$ by the equation of state.  The fluid evolution variables are $\rho_{\star} = \sqrt{\gamma}W\rho_0$, $\tau=\rho_{\star}(hW-1)-P\sqrt{\gamma}$, and $S_i=\rho_{\star}h u_i$.

The hydrodynamic equations are implemented as in earlier papers, with the exception of the treatment of low density regions.  As in Nouri~{\it et al}~\cite{Nouri:2017fvh}, we use an auxiliary entropy evolution variable $\rho S$ to recover primitive variables at points where no solution can be found using the normal conservative variables.  Our density floor is space-dependent:
\begin{equation}
    \rho_{\rm min} = \left[6\times 10^{-5} + \frac{6}{(R/1.47{\rm km})^{1.5} + 1470}\right]\rm{g\  cm}^{-3}
\end{equation}
Also, we no longer impose zero temperature and velocity at very low densities.  Rather, at very low densities, we only set a maximum $T$ of 1 MeV and a maximum Lorentz factor of 2.   The density threshold for applying ceilings is set proportional to $e^{-t/82\rm{ms}}$, so that the density threshold decreases faster than the disk density.  Soon, the threshold is lower than the density nearly everywhere even outside the disk.  In any case, temperatures and Lorentz factors near the ceiling values do not occur except very near the horizon.

\subsubsection{Transport effects}
\label{sec:transport}

Momentum transport is implemented with the ``turbulent mean stress'' method as described in~\cite{Duez:2020lgq}, which is a slight modification of Radice's ``large eddy simulation'' prescription~\cite{Radice:2017}.  Although not strictly equivalent to the relativistic Navier-Stokes equations, it functions as an effective viscosity.  For the strength of effective viscosity, we use an  $\alpha$-viscosity model~\cite{shakura:1973} with $\alpha_{\rm visc}=0.03$.  This corresponds to
effective viscosity
\begin{equation}
\label{eq:eta}
    \eta = \alpha_{\rm visc} P \Omega_K^{-1}
\end{equation}
and mixing length
\begin{equation}
    \ell = \frac{\eta}{c_s(\rho_0+c)}
\end{equation}
where $c\approx 10^2$g cm${}^{-3}$ suppresses effective viscosity at very low densities, which are outside the disk in the polar or outflow regions.

We only apply the $T_{r\phi}$ and $T_{\theta\phi}$ components of the viscous stress, since the $T_{r\theta}$ terms could suppress convective turbulence~\cite{1999MNRAS.303..309I,Fernandez2013,10.1093/mnras/stv009}.

In one run, we add diffusion of composition ($Y_e$) and entropy.  We use the same mixing length, but with a factor to suppress diffusion within roughly a horizon radius of the inner boundary (where it is numerically problematic).  The factor is $1-\exp[-((r-_{\rm in})/r_{\rm in})^3]$, where $r_{\rm in}$ is the inner boundary radius.
See~\cite{Duez:2020lgq} for implementation details.  The turbulent heat flux is set to be

\begin{equation}
    q_i = -\rho_0 T c_s \ell \left[ \partial_i S + \frac{\mu_p + \mu_e - \mu_n}{m_n}\partial_i Y_e\right]
\end{equation}
where $S$ is the entropy per baryon, and $\mu_x$ is the chemical potential of particle $x$.

The evolution equations for the composition and the stress tensor are then modified to be
\begin{eqnarray}
    \nabla_{\mu}(\rho_0 Y_e u^{\mu}) &=& D_j(W\rho_0c_s\ell \partial^j Y_e) + \cdots \\
    \nabla_{\mu}T^{\mu\nu}{}_{\rm gas} &=& -\nabla_{\mu}T^{\mu\nu}{}_{\rm heat} \\
    T^{\mu\nu}_{\rm heat} &=& q^{\mu}u^{\nu} + q^{\nu}u^{\mu}
\end{eqnarray}
where ``$\cdots$'' indicates omitted neutrino sources, and, following a common convention, Latin (spatial) indices run 1--3, Greek (spacetime) indices run 0--3, $\nabla_{\mu}$ is the 4D covariant derivative, and $D_i$ is the spatial covariant derivative.  In general, one could use a different mixing length for each of the transport terms (momentum/viscosity, composition/particle diffusion and heat), but we have chosen to make them all equal.

\subsubsection{Neutrino evolution}
\label{sec:neutrinos}

We evolve neutrino fields using the grey M1 moment closure scheme described in previous SpEC papers~\cite{FoucartM1:2015,Foucart:2016rxm}.  It has been modified in the following ways.

\begin{itemize}
\item We impose a no-inflow constraint on the neutrino number density functions.

\item We only keep charged-current processes for electron-flavor interactions with matter.  We find that including pair production/annihilation processes in particular, with assumptions of Kirchhoff's law and local thermodynamic equilibrium (which are certainly not satisfied when the particles neutrinos are interacting with are other neutrinos in optically thin regions) causes obviously unphysical growth of $Y_e$.

\item Some of our neutrino-matter interaction formulae may behave badly at very low densities, so we turn off neutrino-matter coupling at densities below about $10^{-6}$ the initial maximum density of the disk.  We have tried lowering this cutoff by a factor of 100 and see negligible difference in the composition of the disk and outflow in the first $\sim$ second.

\end{itemize}

After about a second, neutrino luminosities have dropped by five orders of magnitude, and neutrinos are no longer an efficient way of cooling the disk or altering the composition of outflows.  To speed up simulations, we stop evolving neutrinos between 1 an 1.5 seconds after merger.

\subsection{Equation of state}
\label{sec:eos}

We use the Helmholtz equation of state~\cite{Timmes:2000} with its standard treatment of lepton and photon contributions.  This equation of state is coded to use derivatives of the free energy so as to guarantee thermodynamic consistency, which could be important when the advective disk becomes convectively unstable, and modeling the resulting convection requires being able to simulate adiabatic flows at constant composition accurately.  At very high densities ($\rho>10^11$\,g cm${}^{-3}$) and temperatures ($T>10$\,MeV), we switch to analytic expressions for the leptons corresponding to the relativistic limit~\cite{1996ApJS..106..171B}.

The baryon component is a sum of free proton, free neutron, and alpha particle ideal gases in nuclear statistical equilibrium (NSE).  We implement this exactly as in~\cite{Fernandez2013}, using the Saha-like equation
\begin{eqnarray}
    x_n^2x_p^2 &=& \frac{1}{2}x_{\alpha}\left(\frac{n_Q}{n}\right)^3\exp\left[-\frac{Q}{k_BT}\right] \\
    \nonumber
    n_Q &=& \left(\frac{m_n k_B T}{2\pi\hbar^2}\right)^{3/2} \\
    \nonumber
    x_p &=& Y_e - \frac{1}{2}x_{\alpha} \\
    \nonumber
    x_n &=& 1 - Y_e - \frac{1}{2}x_{\alpha}
\end{eqnarray}
where $x_n$, $x_p$, and $x_{\alpha}$ are the mass fraction in free neutrons, free protons, and alpha particles, respectively, $n$ is the number density of nucleons, and $Q=28.29566$\,MeV is the binding energy of a helium-4 nucleus.  This nuclear binding energy contributes $-x_{\alpha}Q/(4 m_n)$ to the total specific internal energy. 

The above prescription automatically incorporates the energy released by helium recombination, and at all densities considered, the ideal gas approximation for baryons is completely adequate.  However, more sophisticated NSE models find that the equilibrium nuclear state at the low densities of our simulation is actually heavy nuclei, which will affect the binding energy released (albeit not by a large amount) and the number density.  Therefore, we have run two simulations using the low-density baryon component taken from the DD2 equation of state~\cite{Hempel:2011mk}.  We use the isolated baryonic component, available in tabulated form at M. Hempel's website~\cite{hempel_2015}.  Similar to the Helmholtz handling of the leptonic component, we use bicubic interpolation (in $\rho_0$ and $T$) of the free energy and its first derivatives.  We interpolate linearly in $Y_e$.  Second derivatives (needed, e.g. for the sound speed), are found by finite differencing tabular values and smoothing in density.

We must extrapolate to densities and temperatures far below the DD2 table minimum (1660 g cm${}^{-3}$ and 0.1 MeV).  For this, we presume an ideal gas, taking the mean molecular weight (equivalently, the mean mass number of baryonic free particles) and nuclear binding energy per baryon to be equal to the nearest corresponding point on the table--e.g. the minimum density point of the given $T$ and $Y_e$ for points with $\rho_0$ below the table minimum. 

\subsection{Modeling of r-process heating}
\label{sec:rprocess}

Even the above prescription for the EoS is questionable, because it assumes the continued validity of NSE at densities and temperatures far below where this is a safe assumption.  In fact, neither an assumption of fast nuclear reactions (NSE) nor an assumption of very slow reactions (constant isotope abundances) is justified, and the only adequate treatment would be to explicitly include a nuclear reaction network with many isotope abundances as new evolution variables.  In particular, it is expected that r-process nucleosynthesis may have important effects on the thermal evolution of outflows~\cite{10.1111/j.1365-2966.2009.16107.x}.

In one run, we incorporate the effect of r-process heating in an approximate but physically motivated way, following the prescription in~\cite{Foucart:2021ikp}.  We add the following driving term to the evolution of $\rho_{\star}Y_e$, which automatically acts as a heating agent, and an expected partially-compensating non-NSE neutrino energy loss term.

\begin{eqnarray}
    \frac{d(\rho_{\star}Y_e)}{dt} &=& \cdots - \rho_{\star}\frac{Y_e - Y_{e,f}}{t_{\rm rp}} \\
    \label{eq:Qdot}
    \nabla_{\mu}T^{\mu\nu} &=& - \dot{Q}_r u^{\nu} \\
    \label{eq:rcooling}
    \dot{Q}_r &=& -0.0085\rho_0 \frac{Y_e - Y_{e,f}}{t_{\rm rp}}
\end{eqnarray}
where $t_{\rm rp}=1$ second and $Y_{e,f}=$0.38.

In our simulation, these terms are applied everywhere in the fluid.  This is not realistic in the inner disk, since above $T\gtrsim 0.5$\,MeV, NSE is applicable.  However, the accretion time is less than $t_{\rm rp}$ for most of the matter in this region, which automatically limits the effect of the r-process on it, and including it everywhere errors to the side of the greatest possible r-process effect~\footnote{Actually, our code includes density and temperature cutoffs on the r-process terms, but due to a bug they were not applied in the run reported in this paper.}.

For our equation of state, we find that $\left.\frac{\partial h}{\partial Y_e}\right|_{\rho_0,T}=-0.026$ and is indeed very nearly constant for $\rho_0\lesssim 10^9$ g cm${}^{-3}$, $T\lesssim 0.1$ MeV, and $Y_e$ between 0.1 and 0.35 and predicts an energy release per rest energy of 0.007 when $Y_e$ changes from 0.1 to 0.38.  [At $Y_e>0.35$, the slope of $h(Y_e)$ flattens, but the effect of the $0.35<Y_e<0.38$ region on the overall energy release is very small.]  The cooling term in Equation~\ref{eq:rcooling} removes an energy per rest energy of 0.002 for the same $Y_e$ change, so the overall effect is indeed a heating with efficiency 0.005.  Furthermore, there is an amplification effect in the disk, in that r-process heating will increase pressure, which increases the viscous heating rate.

This r-process model should only be regarded as a crude approximation, and our implementation has several foreseeable limitations.  The model has the matter move in NSE to higher $Y_e$, while physically it would not be in NSE.  However, Foucart~{\it et al}~\cite{Foucart:2021ikp} point out that the difference in binding energy from NSE is typically $\sim$10\% of the energy released by the r-process, so the NSE approximation is probably acceptable.

The assumption of NSE is also invoked in the computation of neutrino emission rates in the M1 code.  These will, then, also be only approximately accurate in non-NSE regions.  We note that the neutrino emissions leading to the cooling processes of Equations~\ref{eq:Qdot} and~\ref{eq:rcooling} come from different reactions than those modeled by the M1 emission, so it is appropriate for both to operate simultaneously.

Furthermore, note that~\cite{Foucart:2021ikp} only envisioned these approximate r-process terms acting on the outflow. It is unclear how r-process nucleosynthesis would proceed for material orbiting at low density and low temperature in an accretion disk. For material in tidal tails, Metzger {\it et al}~\cite{10.1111/j.1365-2966.2009.16107.x} find that energy deposition due to r-process nucleosynthesis has a strong dependence on the trajectory of the outflow, with the r-process being strongly suppressed after apocenter, when the density of matter starts increasing. Future studies with a full nuclear reaction network will be required to determine what heating rate is appropriate for matter evolving for multiple seconds in an accretion disk. R-process heating has also been included in a couple of other studies of post-merger dynamics~\cite{Fernandez:2014bra,Wu:2016pnw}.

\section{Results}
\label{sec:results}

All runs use the same initial data, include an $\alpha_{\rm visc}=0.03$ effective viscosity and M1 neutrino transport, and evolve for 10 seconds.  The different simulations are labeled as follows.

\begin{itemize}
    \item {\bf Std}, the ``Standard'' run, uses the Helmholtz equation of state with ideal gas baryon component assuming NSE abundances of $p$, $n$, and $\alpha$ particles.  Effective viscosity is the only transport mechanism explicitly included.
    
    \item {\bf Diff}, the ``Diffusion'' run, is identical to Std except that diffusion of $Y_e$ and $S$ have been added with the same mixing length as used for the viscosity, as described in Section~\ref{sec:transport}.
    
    \item {\bf Nuc}, the ``Heavy Nuclei'' run, is identical to Std except that the baryonic component of the equation of state is taken from DD2 (with extrapolation to low densities as described in Section~\ref{sec:eos}).
    
    \item {\bf rProc}, the ''R-Process'' run, is identical to Nuc but with the source terms for $Y_e$ and the energy representing the effect of r-process nucleosynthesis (and neutrino energy loss), as described in Section~\ref{sec:rprocess}.
    
\end{itemize}

Timesteps of the simulation Diff were found to take a factor of a few longer than those of other simulations.  Therefore, we reduced the resolution of Diff by 28\% during the time from 5.4 seconds to 10 seconds after merger.

\subsection{Disk evolution}

Viscosity transports angular momentum outward, leading to accretion of gas into the black hole and expansion of the disk outward.  Observing the matter flow at a fixed radii in what are initially the outer disk, we see first an outflow and then at later times an inflow, which is the expected behavior for a viscosity-driven accretion disk.  (See Section 5.2 of Frank, King, and Raine~\cite{AccretionPower}.)  Viscous heating in the equator drives vertical convection, visible especially in density snapshots.  We note that convection was not seen before restricting viscosity to $T_{r\phi}$ and $T_{\theta\phi}$ and implementing the Helmholtz equation of state.

\begin{figure}[h]
  \includegraphics[width=8cm]{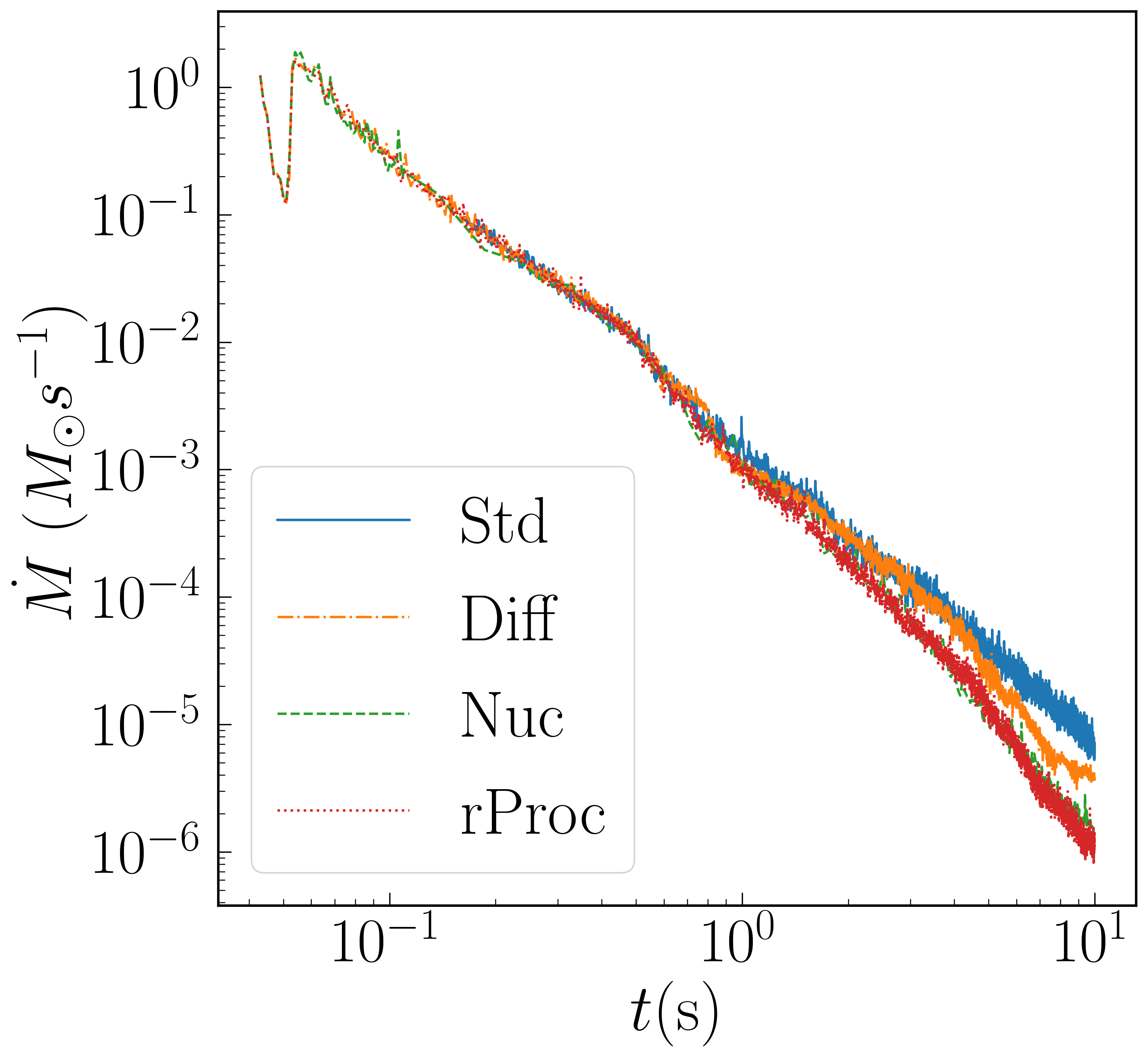}
  \includegraphics[width=8cm]{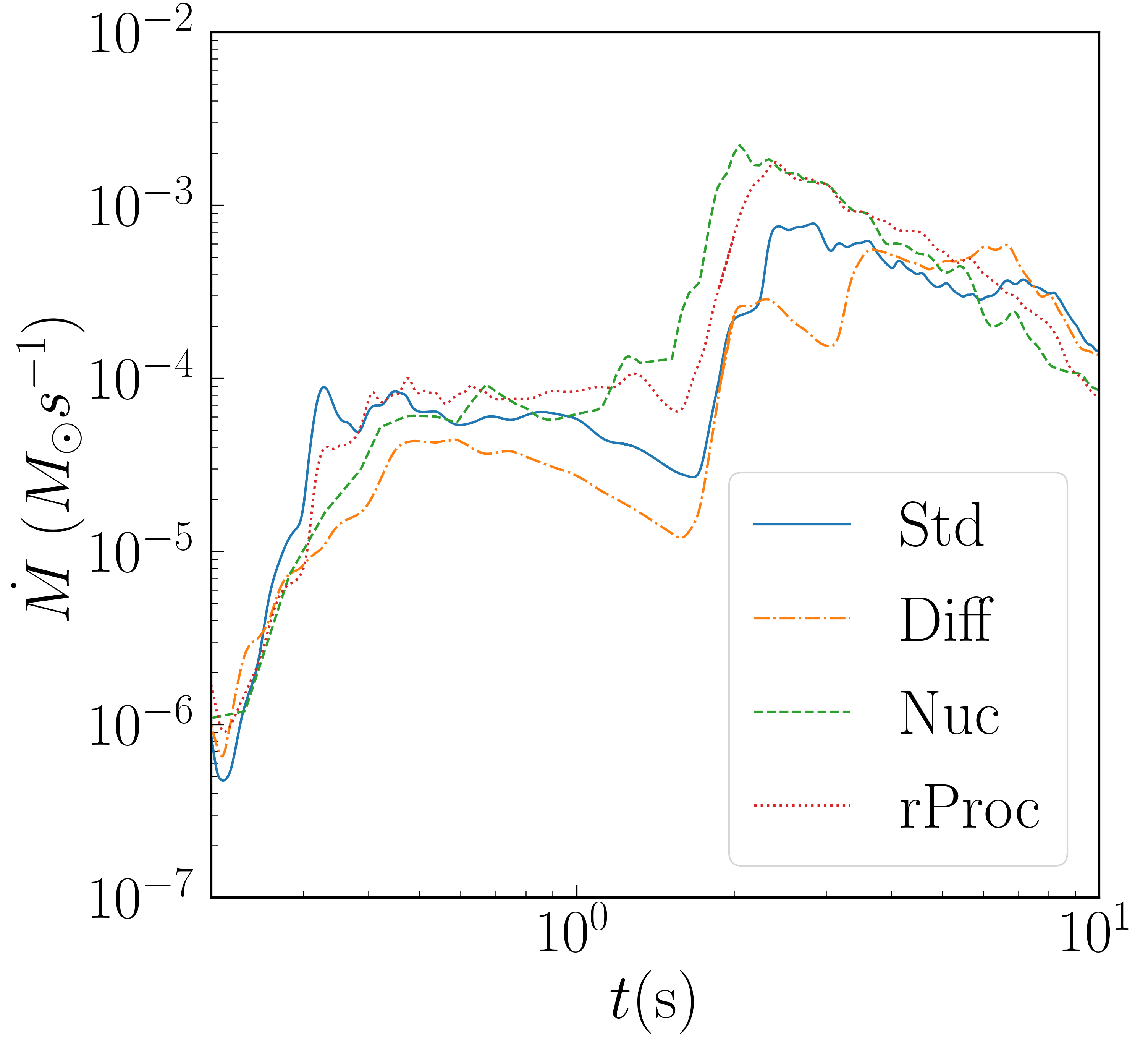}
  \caption
  {Left:  rate of accretion onto the black hole.  Right:  Rate of outflow through radius $r=$39,000\,km.}
  \label{fig:accretion}
\end{figure}

Figure~\ref{fig:accretion} shows the rate of advection of baryonic mass into the black hole and out of a large observation radius, i.e. the accretion rate $\dot{M}_{\rm acc}$ and outflow rate $\dot{M}_{\rm out}$.  The initial state is not in equilibrium, and the initial $\dot{M}_{\rm acc}$ is already non-zero even before viscosity is turned on.  There is a dip at early times as this initial accretion subsides and before viscosity-driven evolution begins.  Once viscosity is activated, the accretion rate jumps to $\sim M_{\odot}$ s${}^{-1}$ and subsequently follows an approximate power law in time, $\dot{M}_{\rm acc}\propto t^{-n}$ where $n=2.1\pm 0.1$ for run Std.  This is similar to the $n=1.9$ found for viscous Newtonian hydrodynamics by Fernandez~{\it et al}~\cite{10.1093/mnras/sty2932} for a slightly different system.  It is observably steeper than the $n=1.5$ found for viscous relativistic hydrodynamics by Fujibayashi~{\it et al}~\cite{Fujibayashi:2020jfr} for slightly different systems.  (This paper used isentropic equilibrium initial data with analytically specified $Y_e(\rho_0)$ and $hu_{\phi}(\Omega)$ initial functions, with $\Omega$ the angular velocity.  Their black hole masses closest to ours are 6 and 10\,$M_{\odot}$.)

\begin{figure}[h]
  \includegraphics[width=8cm]{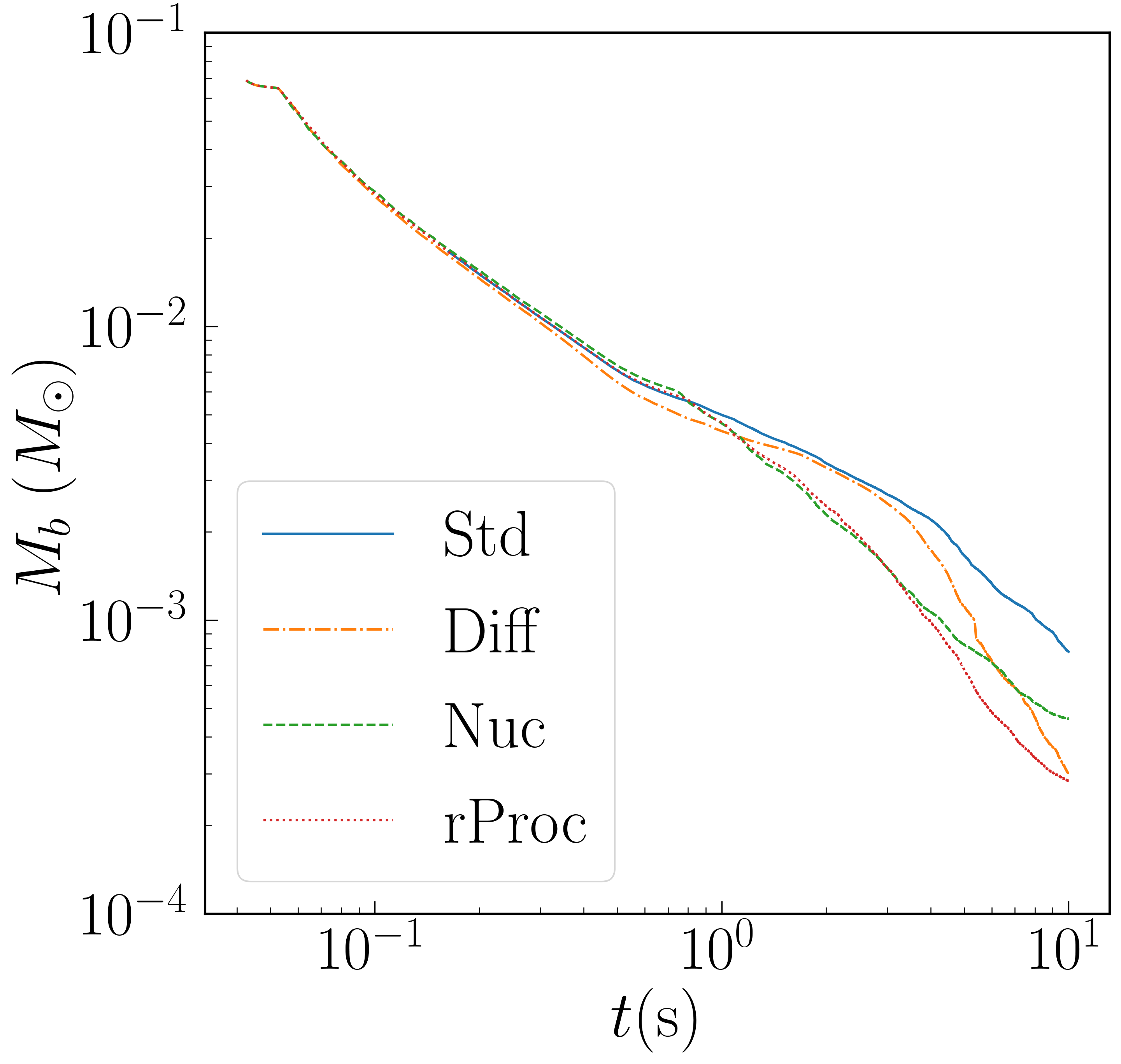}
  \includegraphics[width=8cm]{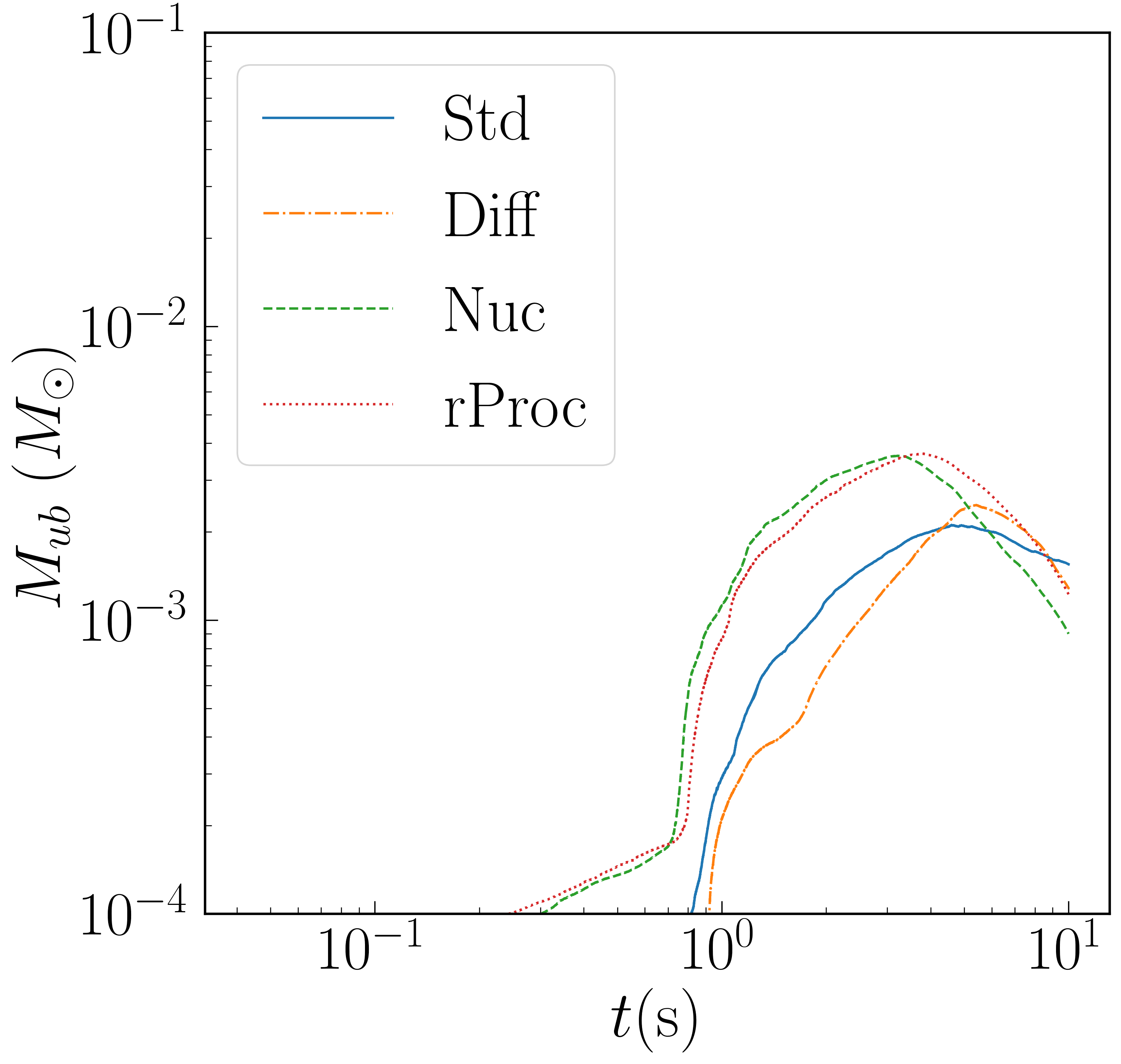}
  \caption
  {Overall evolution of mass on the grid.  Left:  bound baryonic mass on the grid.  Right:  Unbound mass on the grid.}
  \label{fig:M}
\end{figure}

The other runs have the same $\dot{M}_{\rm acc}$ for the first second but show noticeably slower accretion thereafter.  The reason appears to be that the inner disk is more depleted at $t>1$s for other runs as compared to Std.  Simulations with heavy nuclei (Nuc and rProc) unbind more material in outflow after about a second, which can clearly be seen in the plot of $\dot{M}_{\rm out}$.  It can be seen even more directly in Figure~\ref{fig:M}, which plots the total bound and unbound mass on the grid as a function of time (with unbound defined as $u_t<-1$).  The accretion rate for Diff drops well below that of Std at later times (about 5 seconds post-merger).  Diff does not have greatly more unbound mass than Std, so the explanation is different.  The last 5 seconds of Diff was run at a lower resolution, so its late-time evolution does have larger error than the other runs.  However, the divergent evolution from Std is quite believable when one looks at disk profiles which show density, entropy, and $Y_e$ growing distinct for the Diff run well before the switch to low resolution.

\begin{figure}[h]
  \includegraphics[width=8cm]{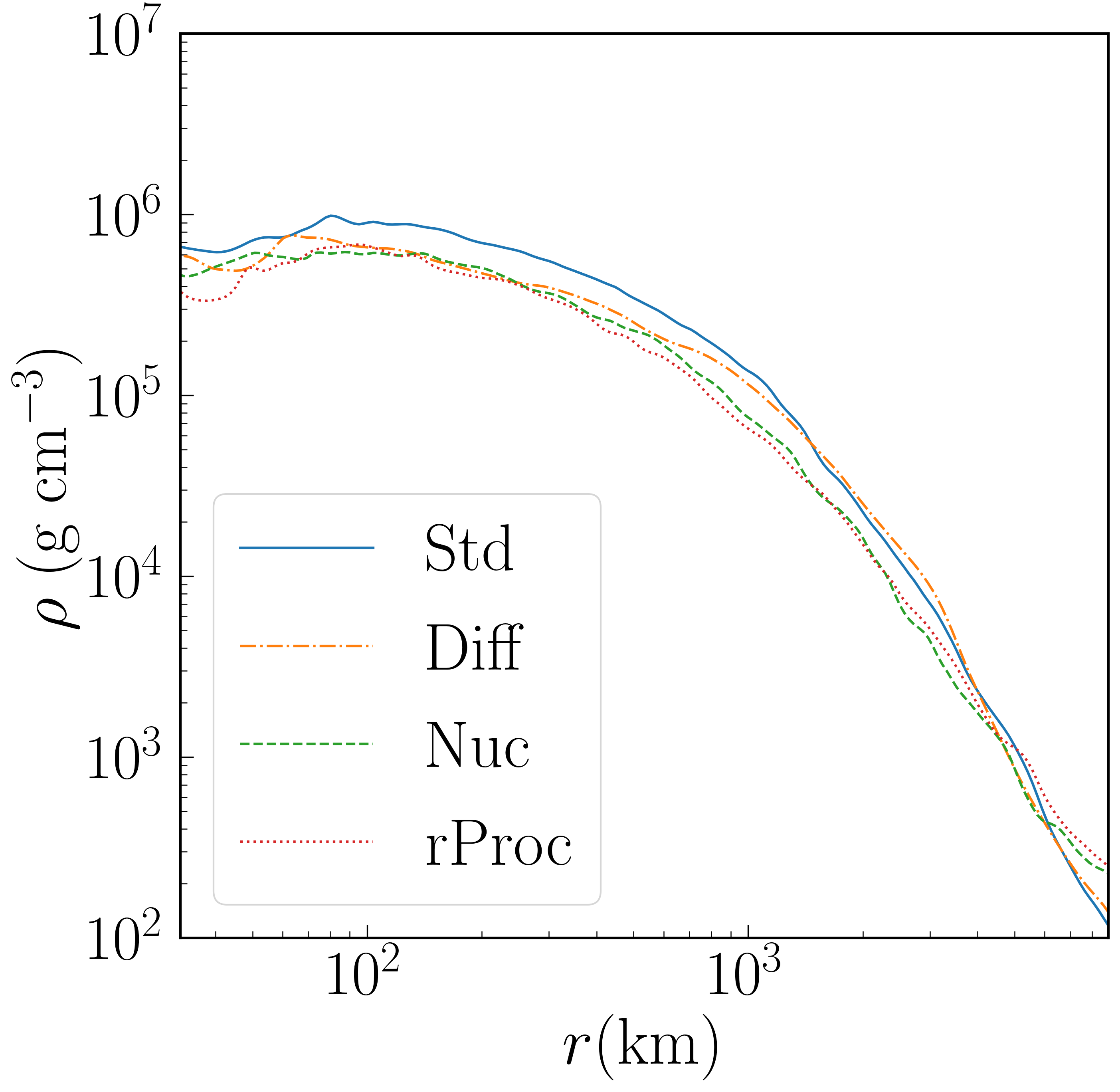}
  \includegraphics[width=8cm]{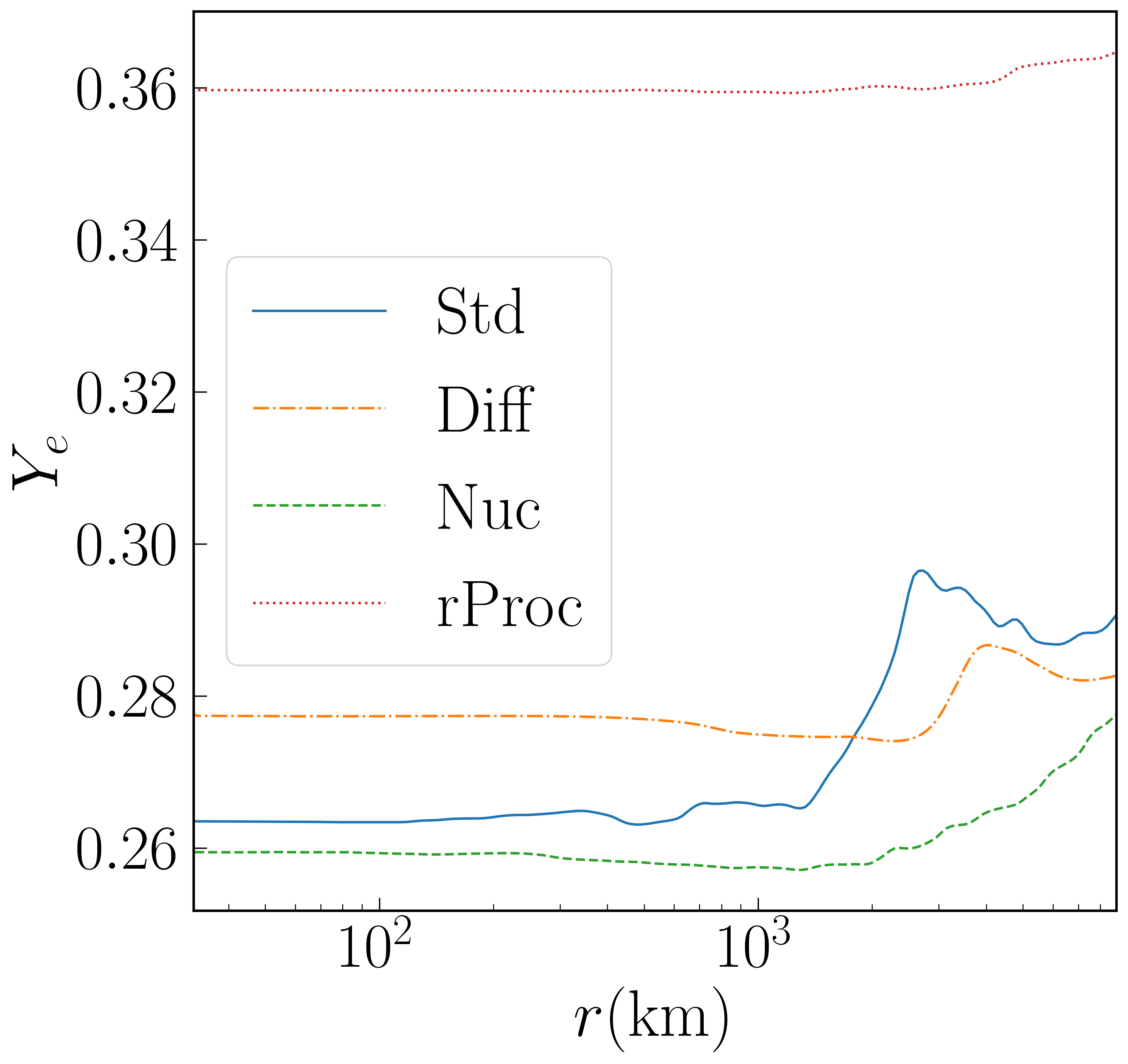} \\
  \includegraphics[width=8cm]{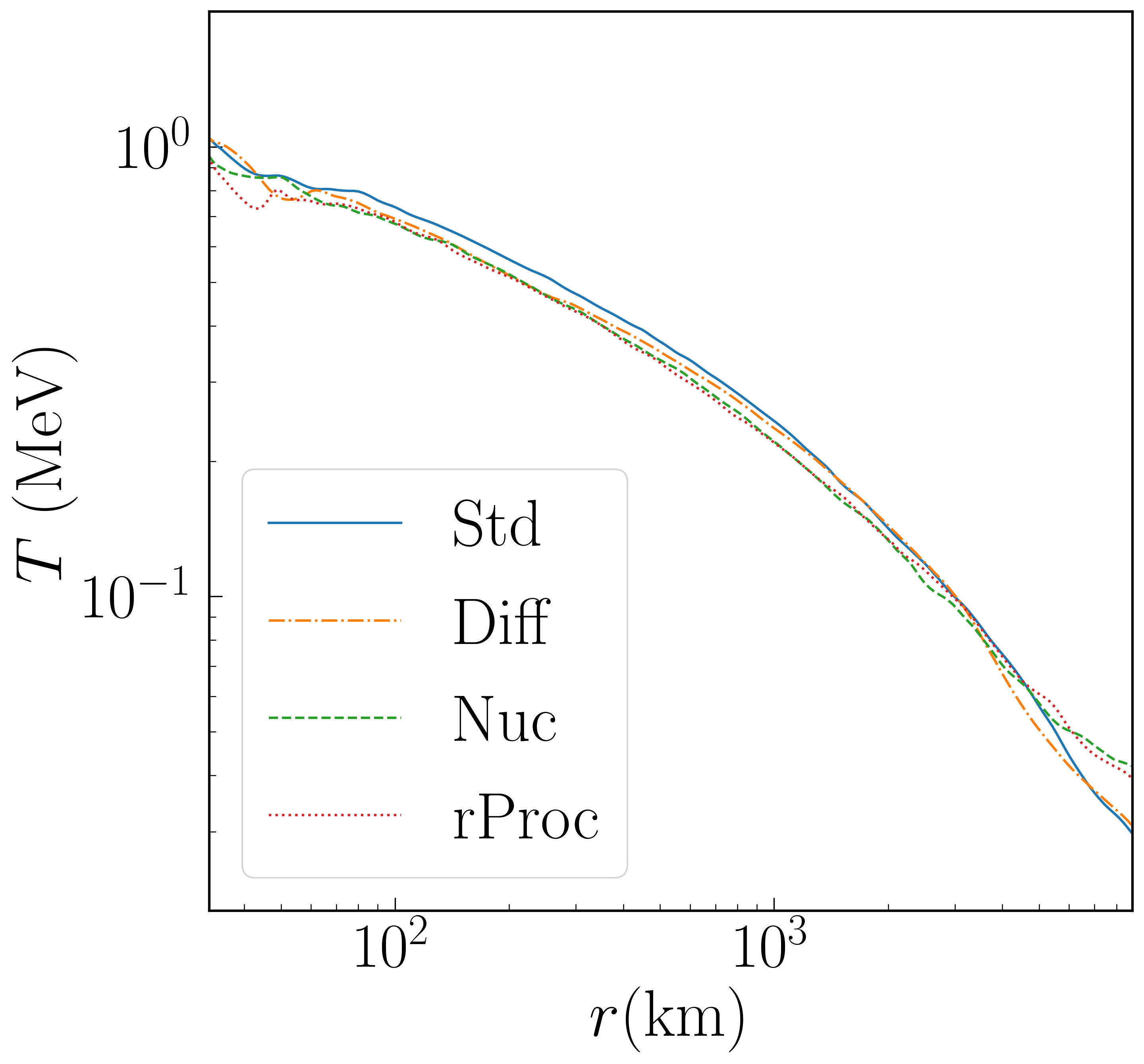}
  \includegraphics[width=8cm]{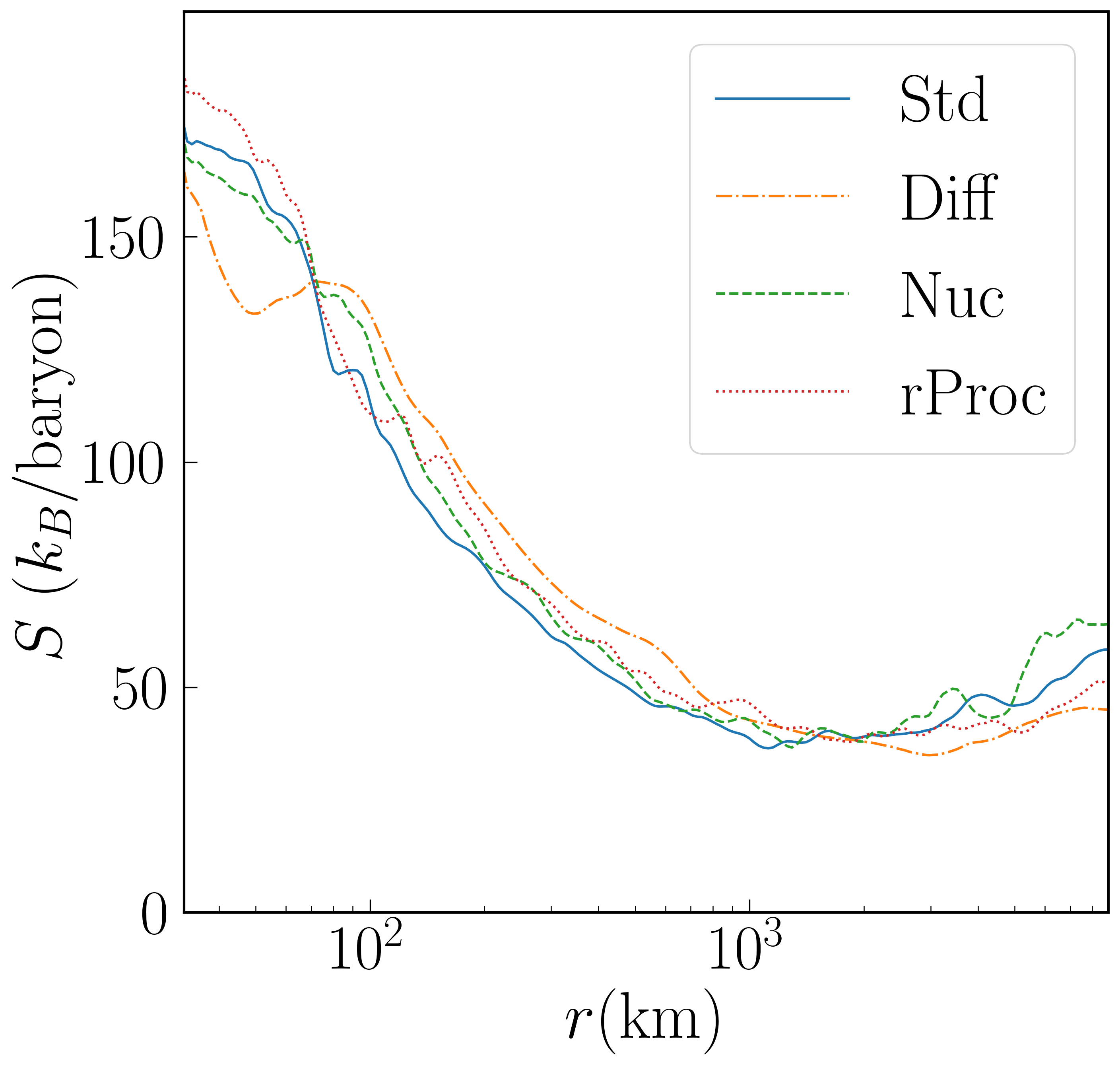}
  \caption
  {Spherically-averaged radial profiles of the disk region at 2 seconds after merger.  Composition, temperature, and specific entropy averages are density-weighted.}
  \label{fig:profiles}
\end{figure}

These profiles are shown in Figure~\ref{fig:profiles} for a time 2 seconds after merger.  We plot the inner 9000\,km.  While profiles appear continuous, the fraction of matter that is unbound raises from 0.1 at $r\sim 4000$\,km to 0.9 at around $r\sim 9000$\,km.  There is no unbound material inside $r<2300$\,km.  The density of the disk is indeed seen to be higher for Std than for other runs.

Most of the profiles change significantly with time, the exception being the temperature profile, which is roughly constant because it is set by the requirement of dynamical equilibrium for a thick disk:  $k_BT\sim Mm_p/r$.  Note, therefore, that since the density of the disk is continually decreasing, its specific entropy is continually increasing.  It is important to be clear about what processes do and do not affect the specific entropy.  Adiabatic compression and recombination of nuclei in NSE are reversible effects which increase temperature but not entropy.  The source of entropy growth in these simulations is viscous heating.  (The only other conceivable sources, shock heating and neutrino absorption, are comparatively unimportant.)  During the first half-second or so, this heating is partly offset by neutrino cooling, but afterwards the disk enters and advective state, and the observed continual entropy growth is expected.  It should be recognized, however, that recombination in NSE can indirectly affect the entropy, because increasing the temperature increases the viscosity $\eta$ (cf. Equation~\ref{eq:eta}).  This provides one means whereby Std and Nuc, which differ primarily in the latter's inclusion of recombination to heavy nuclei, will have distinct entropy evolutions.

By 2 seconds, the $Y_e$ source term in rProc has increased $Y_e$ to near its target value everywhere.  The r-processing of the innermost disk is mostly the result of the unphysical inclusion of the r-process in those high temperature regions, but it should also be remembered that the gas near the black hole at $t=2$ seconds is not the gas that was initially there, but gas that was initially farther away at lower temperature that has accreted inward.  The viscous timescale $\tau_{\rm visc}\sim \alpha^{-1}(r/H)^2\sqrt{r^3/M}$ is 0.33\,s for $r\approx 100$\,km and 2\,s for $r\approx 340$\,km, around where $T=0.5$\,MeV  (although the gas near the black hole at 2\,s would have presumably spent most of its time inside this radius).


For Diff, the diffusion effects have succeeded in partially leveling $Y_e$ (increasing it in the inner region) and $S$ (decreasing it in the very hot inner region).  Some care must be taken with this interpretation, though, since the effective heat conduction alters the vertical entropy gradient and partially suppresses convection~\cite{Duez:2020lgq}.  Thus, Std and Diff are really two models of mixing.

\begin{figure}[h]
  \includegraphics[width=5.1cm]{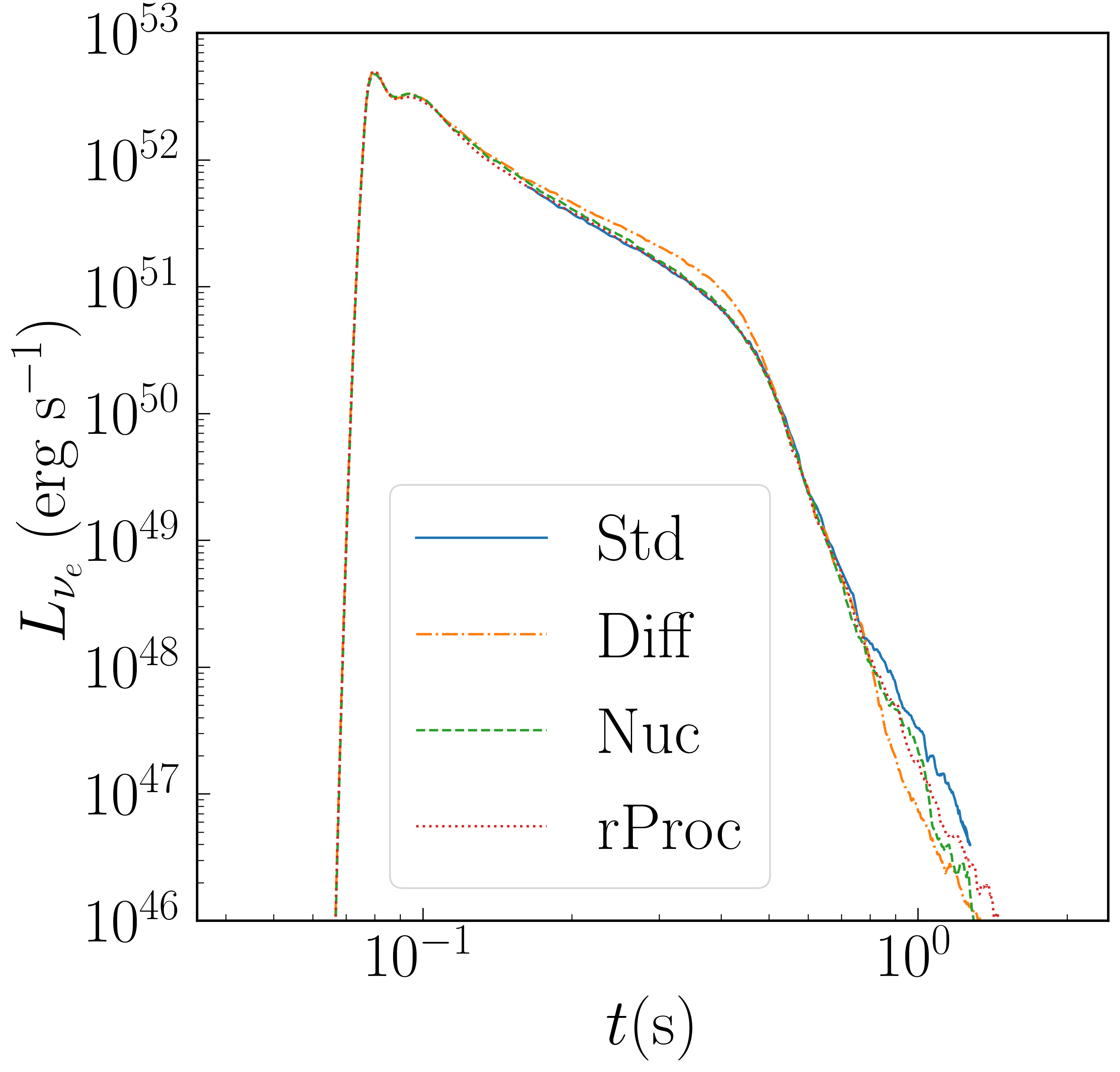}
  \includegraphics[width=5.1cm]{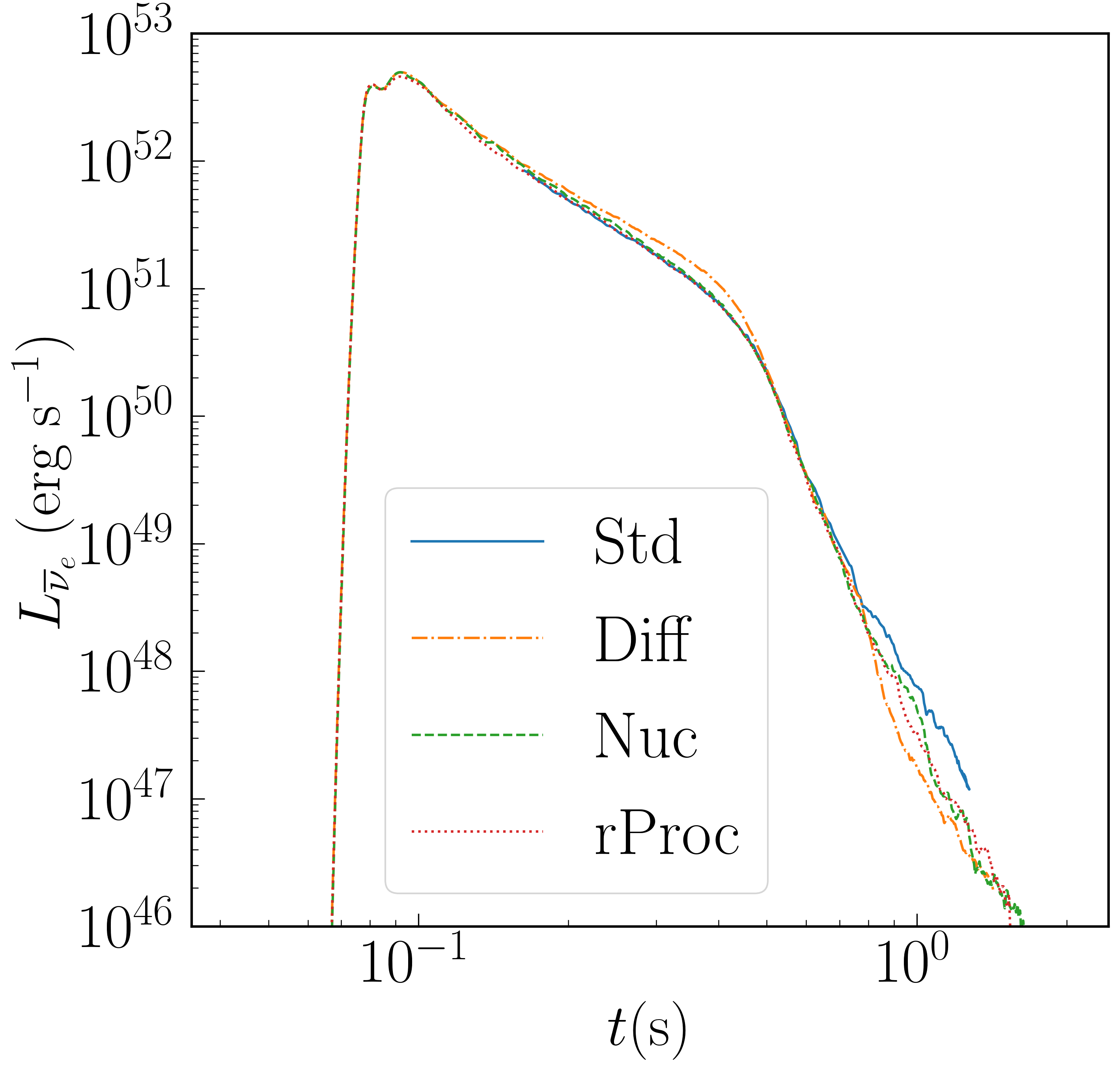}
  \includegraphics[width=5.1cm]{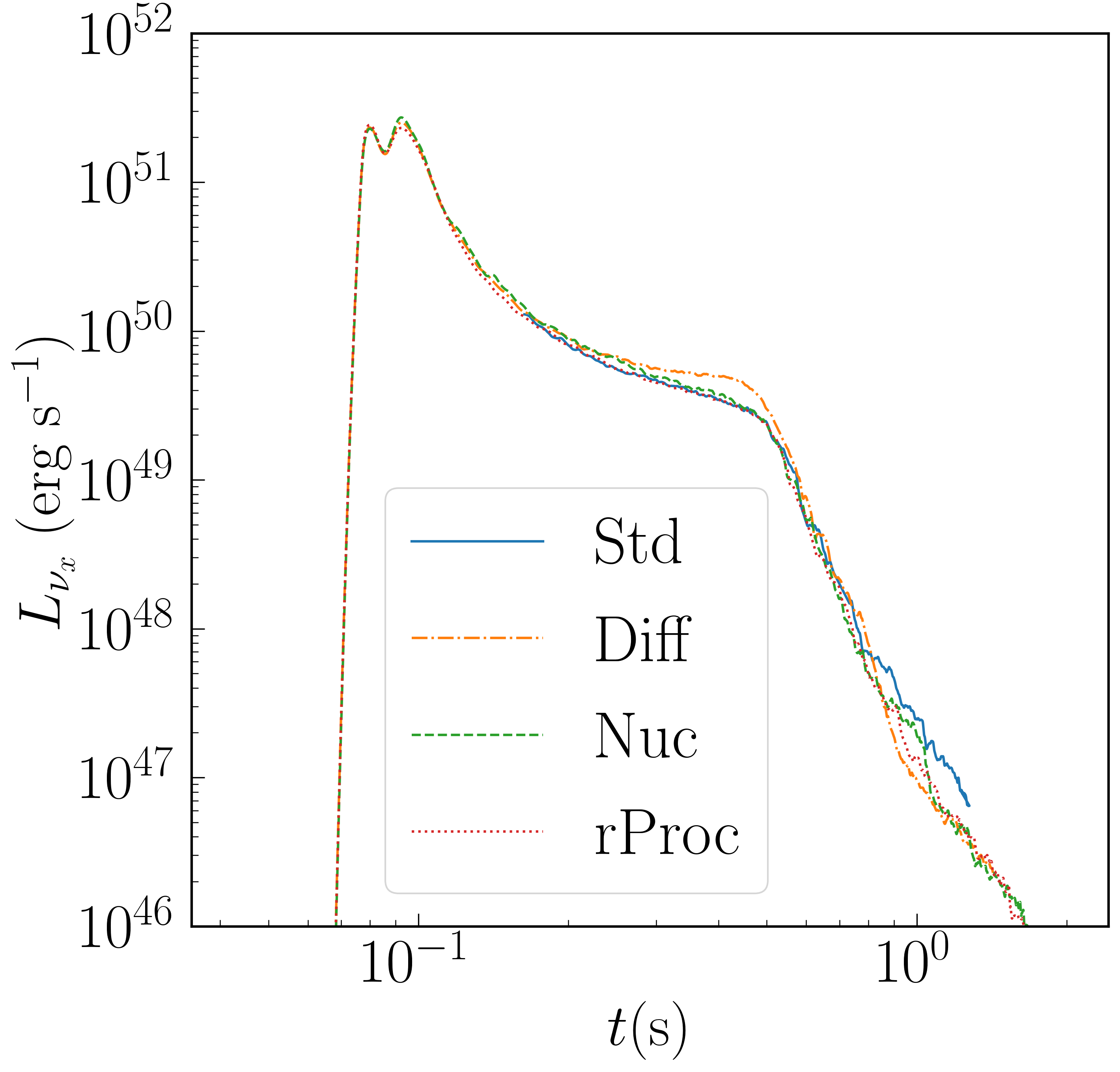}
  \caption
  {Neutrino luminosity passing through a radius of 10,000km.  The jump upward from zero reflects the time for neutrinos to travel to the extraction radius.  Plots stop at around 1.5 seconds after merger, when the evolution of neutrino fields was stopped.}
  \label{fig:neutrinos}
\end{figure}

Neutrino luminosities are shown in Figure~\ref{fig:neutrinos}.  We plot three neutrino species:  electron neutrinos, electron anti-neutrinos, and all other flavors and their anti-particles (labeled ``x'').  There are no noticeable differences between cases.  Very early in the simulation, the disk becomes completely optically thin.  Neutrino cooling remains important for the first few hundred milliseconds.  Its effect on the disk is most easily seen in the mass-averaged $Y_e$, which climbs during this time due to greater emission of $\nu_{\overline{e}}$ than $\nu_e$.  After $t=0.5$\,s, the luminosities begin to fall off more steeply, and the disk becomes radiatively inefficient.  After one second, neutrino emission is negligible, the average $Y_e$ on the grid is nearly constant, and evolution of the neutrino fields is discontinued.

\begin{figure}[h]
  \includegraphics[width=8cm]{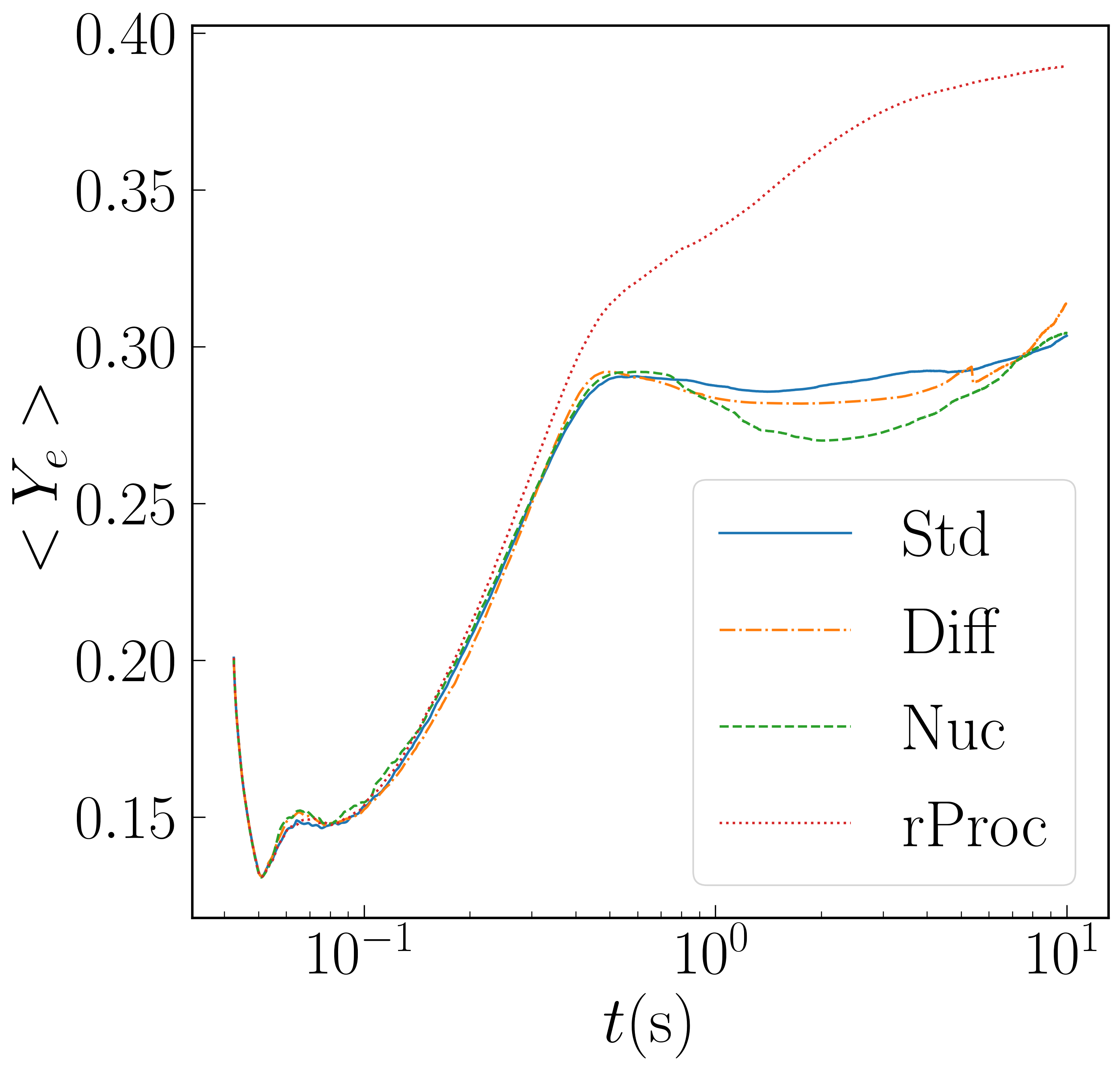} \ \ \ 
  \includegraphics[width=7.7cm]{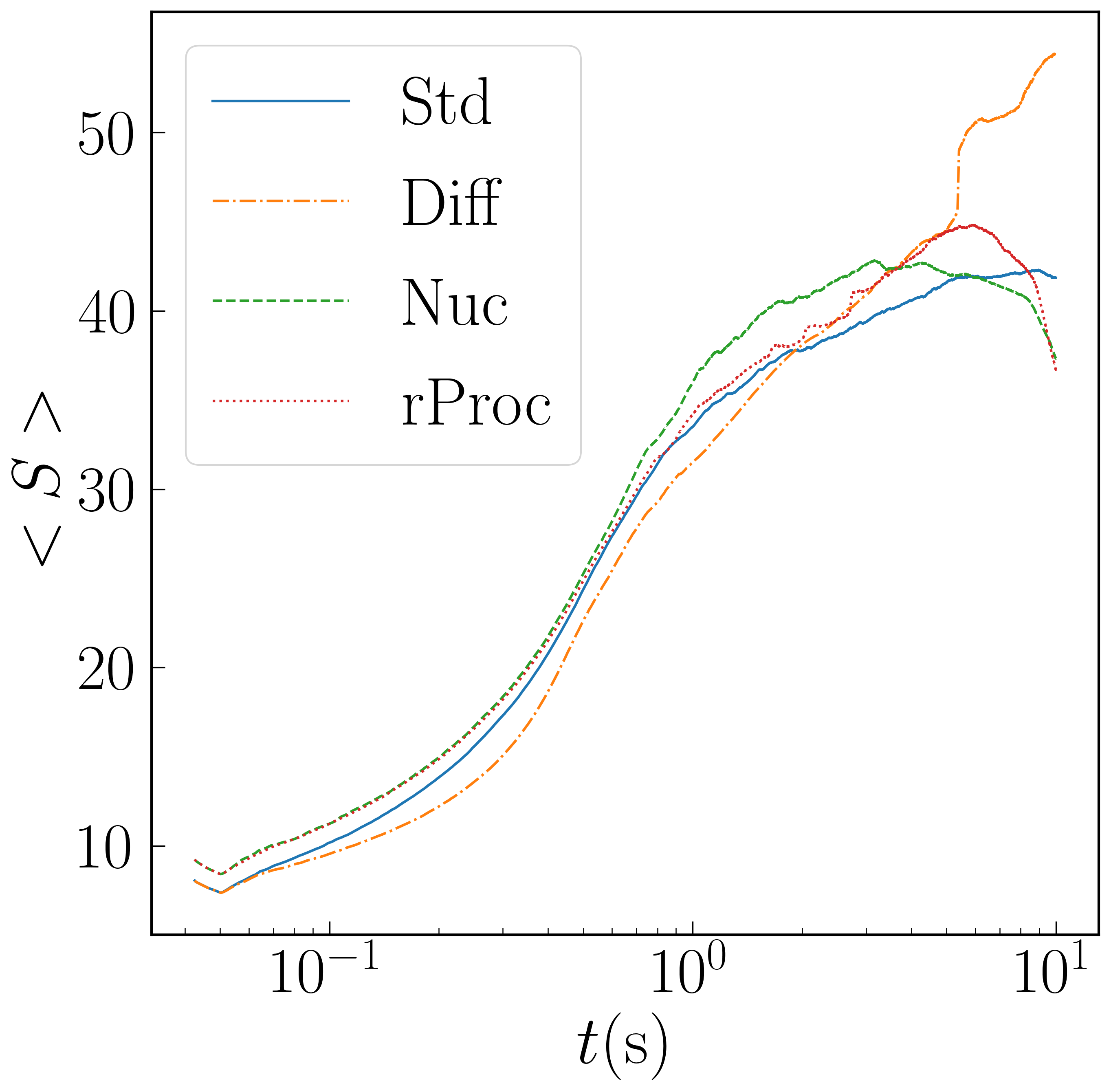}
  \caption
  {Mass-weighted average of $Y_e$ and specific entropy $S$ for bound matter (defined as $-u_t-1<0$).  A glitch can be seen in run Diff when resolution is lowered, the result not of interpolation error but of removing a few layers of high-density, low-entropy gridpoints to move the inner boundary slightly outward.}
  \label{fig:Ye-vs-t}
\end{figure}

In Figure~\ref{fig:Ye-vs-t} we plot the mass-weighted average of $Y_e$ of bound matter on the grid, denoted $\langle Y_e\rangle$.  During the first second, evolution of $\langle Y_e\rangle$ is dominated by neutrino processes (mostly emission).  After the first second, neutrino effects become negligible, and $Y_e$ of a fluid element can only change by the r-process (for run rProc).  $\langle Y_e\rangle$ can continue to change by accretion and outflow, e.g. it will decrease if gas of above average $Y_e$ falls into the black hole or becomes unbound.  In run Nuc, $\langle Y_e\rangle$ drops and then rises mainly because first higher $Y_e$ matter and then lower $Y_e$ matter is marked as unbound.  The average $Y_e$ for all matter on the grid is more nearly constant during the fall and rise times of $\langle Y_e\rangle$ for bound matter, although the average $Y_e$ of all matter on the grid drops between about 3 and 5 seconds when higher-$Y_e$ outflow leaves the grid.

Disks with efficient neutrino cooling are subject to a regulating mechanism that keeps $Y_e\approx 0.1$~\cite{Beloborodov:2002af,Siegel2017}, but this will not operate for this disk for most of its evolution.  This figure also shows the mass-weighted average entropy of bound matter, with increases as the disk density decreases while the temperature at each radius changes much less.

\subsection{Outflow properties}

Given an observation radius $r_{\rm obs}$, we compute outflows as follows.  First, we only consider unbound matter.  We note that matter at $r>4\times 10^4$\,km is all unbound by the definition $u_t<-1$ and hence also by the definition $h u_t/h(T=0)<-1$.  The definition is more important in the interior, and we use the stricter condition $u_t<-1$.  The ejected mass measured at radius $r_{\rm obs}$ at time $t$ is defined to be the sum of the unbound mass that has passed through $r_{\rm obs}$ by time $t$ and the unbound mass interior to $r_{\rm obs}$ at time $t$.
\begin{equation}
    M_{\rm out} = \int^t dt' \int_{r=r_{\rm obs}} d\vec{S}\cdot \vec{v}\rho_{\star} + \int_{r<r_{\rm obs}} d^3x \rho_{\star}f_{\rm ub}
    \label{eq:Mout}
\end{equation}
where $f_{\rm ub}=1$ if the fluid at a point is unbound and $=0$ if it is bound.  By $t=10$\,s, the second term is negligible for $r_{\rm obs}<10^4$\,km.

The mass-averaged value of intensive variable $X$ is defined as
\begin{equation}
    \langle X\rangle = \int^t dt' \int_{r=r_{\rm obs}} d\vec{S}\cdot \vec{v}\rho_{\star}X + \int_{r<r_{\rm obs}} d^3x \rho_{\star}f_{\rm ub}X
\end{equation}

One important intensive variable is the asymptotic velocity $V_{\infty}$, a proxy for the asymptotic Lorentz factor, which, continuing to use $u_t$ as an energy proxy, we compute as
\begin{equation}
    V_{\infty} = \sqrt{1 - (u_t)^{-2}}
\end{equation}

Finally, we compute the mass distribution of $V_{\infty}$ and $Y_e$ in the outflow.  This can be defined in two ways.  Consider intensive quantity $X$, whose range is divided into bins with equal-width $\Delta X$.  Ejecta histograms often display the amount of outflow mass in each bin, i.e. for the $i$-th bin, $\Delta M_X(X_i+\Delta X/2)$ is defined to be the amount of outflow mass with $X$ between $X_i$ and $X_i + \Delta X$.  A plot of $\Delta M_X$ shows at a glance the amount of mass in each bin, but of course this number depends on the chosen $\Delta X$.  This dependence can be removed by defining the distribution $M_X = \Delta M_X / \Delta X$.  Below, we plot $\Delta M_X$ but provide $\Delta X$ so that readers can convert to $M_X$ if they wish.

Ejecta masses and mass-averaged quantities are given in Table~\ref{tab:ejecta}.  Considering first the ejecta mass itself, we do see noticeable effects of our modeling choices.  In particular, 60\% more matter is unbound from the disk if one uses the nuclear equation of state component of Nuc and rProc.  R-process heating itself seems to have a much more modest effect.   This is consistent with Just~{\it et al}~\cite{~\cite{10.1093/mnras/stv009}}, whose simulations with an r-process heating prescription show little difference in ejected mass.  For comparison, Fernandez~{\it et al}~\cite{Fernandez:2016sbf} evolving this same case (their model Fdisk), found an ejected mass of $0.0048 M_{\odot}$.  Our model with the same equation of state (run Std) has $M_{\rm out}=0.0054 M_{\odot}$, which is quite close given the differences in the treatment of gravity and neutrinos.

\begin{table}
        \begin{center}
        \caption[Ejecta properties]{ 
                Ejecta properties for each run:  the total ejected baryonic mass $M_{\rm 0,ej}$ and mass-weighted averages of $Y_e$, entropy per baryon $S$, and asymptotic velocity $V_{\infty}$, with the latter calculated at $10\times 10^3$\,km and $88\times 10^3$\,km.
                }
        \label{tab:ejecta}
\begin{tabular}{l ccccc }
        \toprule \toprule
        Model & $M_{\rm out}/10^{-3} M_\odot$ & $\langle Y_e \rangle$ &
        $\langle S \rangle/k_B$ & 
        $\langle V_{\infty}\rangle_{10}/0.01\,c$ & $\langle V_{\infty}\rangle_{88}$/0.01\,c \\
        \midrule
        Std   & 5.4 & 0.30 & 23 & 6.1 & 7.4\\
        Diff  & 3.6 & 0.30 & 23 & 4.1 & 7.2\\
        Nuc   & 8.6 & 0.30 & 33 & 5.9 & 9.7\\
        rProc & 8.3 & 0.39 & 22 & 4.9 & 7.6\\
        \bottomrule \bottomrule
\end{tabular}
        \end{center}
\end{table}

We see a greater difference from the Newtonian leakage simulation~\cite{Fernandez:2016sbf} in $\langle V_{\infty}\rangle$.   The average $V_{\infty}$ measured at $r_{\rm ob}=10^4$\,km is 0.039\,$c$ in the Newtonian simulation but 0.054\,$c$ in Std.  Might this be a difference of the relativistic treatment?  Relativistic MHD simulations (including 2D MHD dynamo simulations) do produce much faster average outflow velocities, of around 0.1\,$c$~\cite{10.1093/mnras/sty2932,Shibata:2021xmo}, but this is largely because of an early-time high-velocity MHD outflow, which will also be absent in the relativistic viscous simulations reported here.  The late-time ``thermal'' disk outflow in MHD simulations is closer in speed to that of their viscous counterparts.  2D relativistic viscous hydrodynamic simulations of tori resembling post-merger configurations have been carried out by Fujibayashi~{\it et al}~\cite{PhysRevD.101.083029,Fujibayashi:2020jfr}.  Their initial conditions are artificial (exact equilibria, constant entropy), but they vary black hole and disk mass, as well as initial angular velocity profile.  They find ejecta velocities in the range 0.05-0.1\,$c$.  The closest configuration in that paper to the one studied here was perhaps model M06L05, with $M_{\rm BH}=6 M_{\odot}$, $M_{\rm disk}=0.1M_{\odot}$, $a_{\rm BH}/M_{\rm BH}=0.8$.  At simulation end, this model has $M_{\rm out}=0.0087M_{\odot}$, $\langle Y_e\rangle=0.379$, $\langle S\rangle=27.5k_B$, $\langle V_{\infty}\rangle\approx 0.07c$.  Their equation of state is DD2 at high densities, so run Nuc is presumably the best comparison.  Given that our initial states are still somewhat different, their $M_{\rm out}$, $\langle S\rangle$, and $\langle V_{\infty}\rangle$ are reasonably close to ours.  Overall, Fujibayashi~{\it et al}'s results do indicate that ejecta velocities $\gtrsim 0.05c$ are common in relativistic viscous evolutions of post-merger scenarios, and our having simulated an exact case with Newtonian counterpart confirms that the difference is probably due to the relativistic treatment.  In fact, our average asymptotic velocities measured at 10,000\,km underestimate the true asymptotic velocity, because measurement at further radii demonstrates that the outflow is still gaining kinetic energy beyond that point, a point to which we will return.

Our $\langle Y_e\rangle \approx 0.3$ is somewhat smaller than that found by Fernandez~{\it et al}~\cite{Fernandez:2016sbf} for this case,  $\langle Y_e\rangle=0.35$, and that Fujibayashi~{\it et al}~\cite{Fujibayashi:2020jfr} found for their model M06L05, $\langle Y_e\rangle=0.379$.  It is close to the values of $\langle Y_e\rangle$ that Fujibayashi~{\it et al}~\cite{PhysRevD.101.083029} and Just~{\it et al}~\cite{10.1093/mnras/stv009} found for outflows from systems with lower mass black holes.  It is difficult to identify the cause of this difference.  All simulations except that of Fernandez~{\it et al} involve somewhat different initial states, and neutrino transport is handled at least slightly differently by all codes.  One concern about the simulations in this paper is the use of lower density cutoffs on neutrino interactions and on turning off neutrino effects altogether after one second.  To check the effect of these simplifications, we reran Nuc and Diff for 2 seconds with density cutoffs a factor of 100 lower.  We note that the mass-averaged $Y_e$ for unbound mass on the grid settles by about one second.  We find that the change in outflow composition is completely negligible.  It remains possible that the cause of the differences is in the transport and leakage algorithms themselves, in which case the correct outflow composition for viscous hydrodynamic evolution will only be determined by a full neutrino transport scheme. 

\begin{figure}[ht]
  \includegraphics[width=7.8cm]{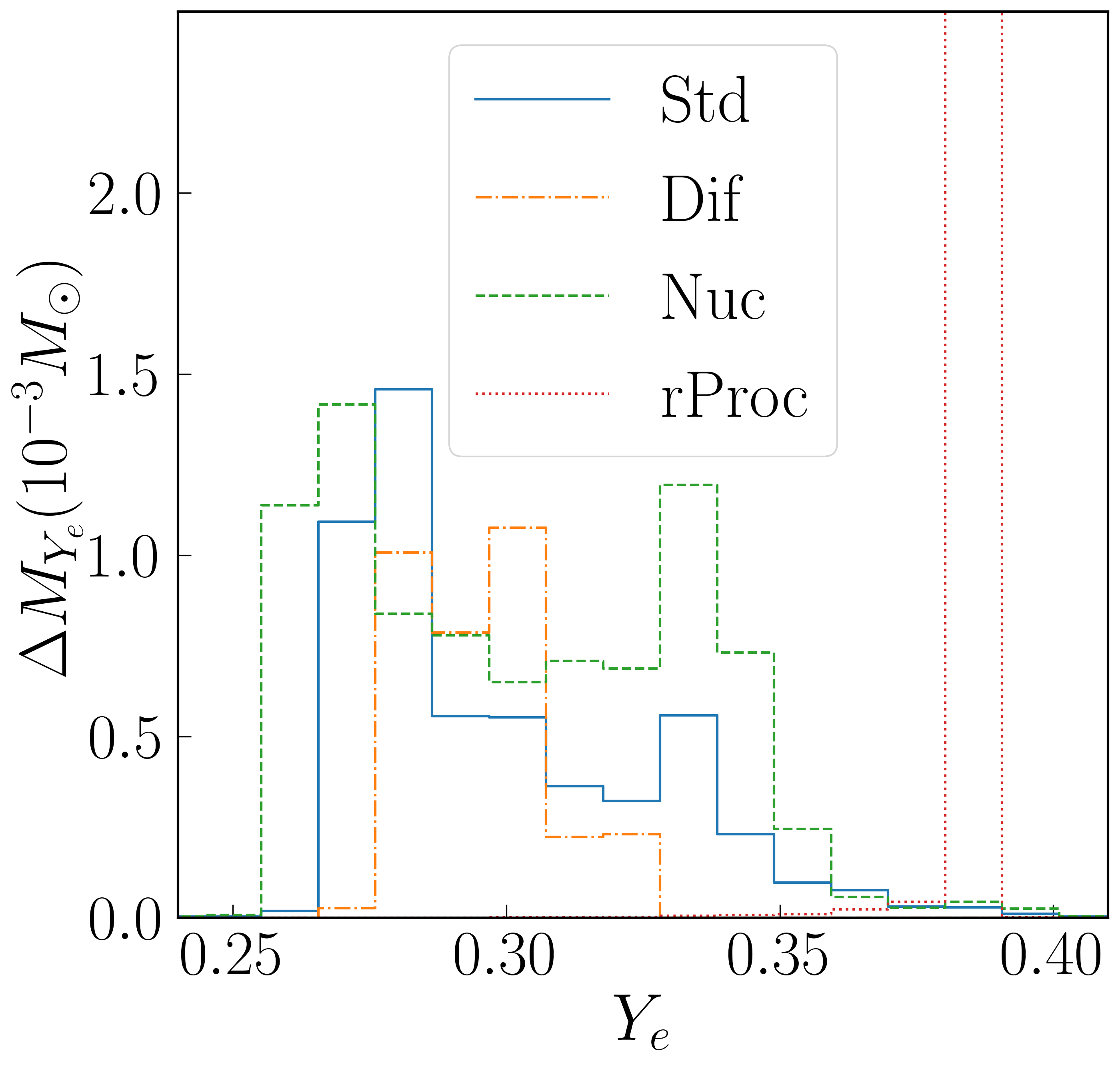}
  \includegraphics[width=7.8cm]{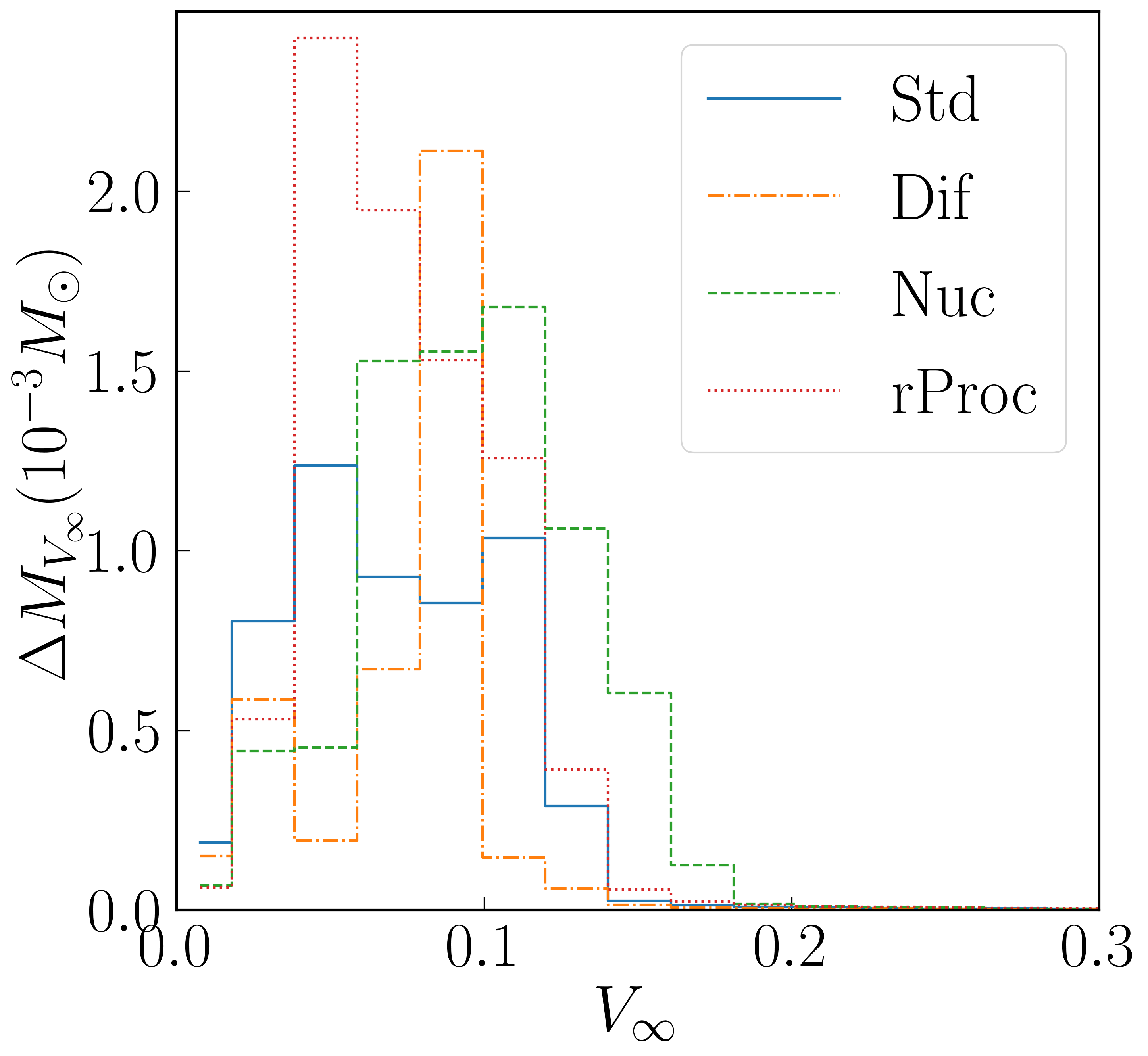}
  \caption
  {Distributions of $Y_e$ and $V_{\infty}$
  of unbound outflow inside or having passed through an extraction radius of 88,000\,km at 10 seconds post-merger.  The bin widths are $\Delta Y_e=0.01$, $\Delta V_{\infty}=0.02$.}
  \label{fig:hist}
\end{figure}

Distributions of $Y_e$ and $V_{\infty}$ are shown in Figure~\ref{fig:hist}.  Particle diffusion effects seem to be able to modestly reduce the spread of $Y_e$ in the Diff run, as might be expected.  In fact, diffusion terms exert a less obvious but stronger effect on the $V_{\infty}$ distribution, which is bimodal but has a very sharp peak at around 0.09\,$c$.

\begin{figure}[h]
  \includegraphics[width=8cm]{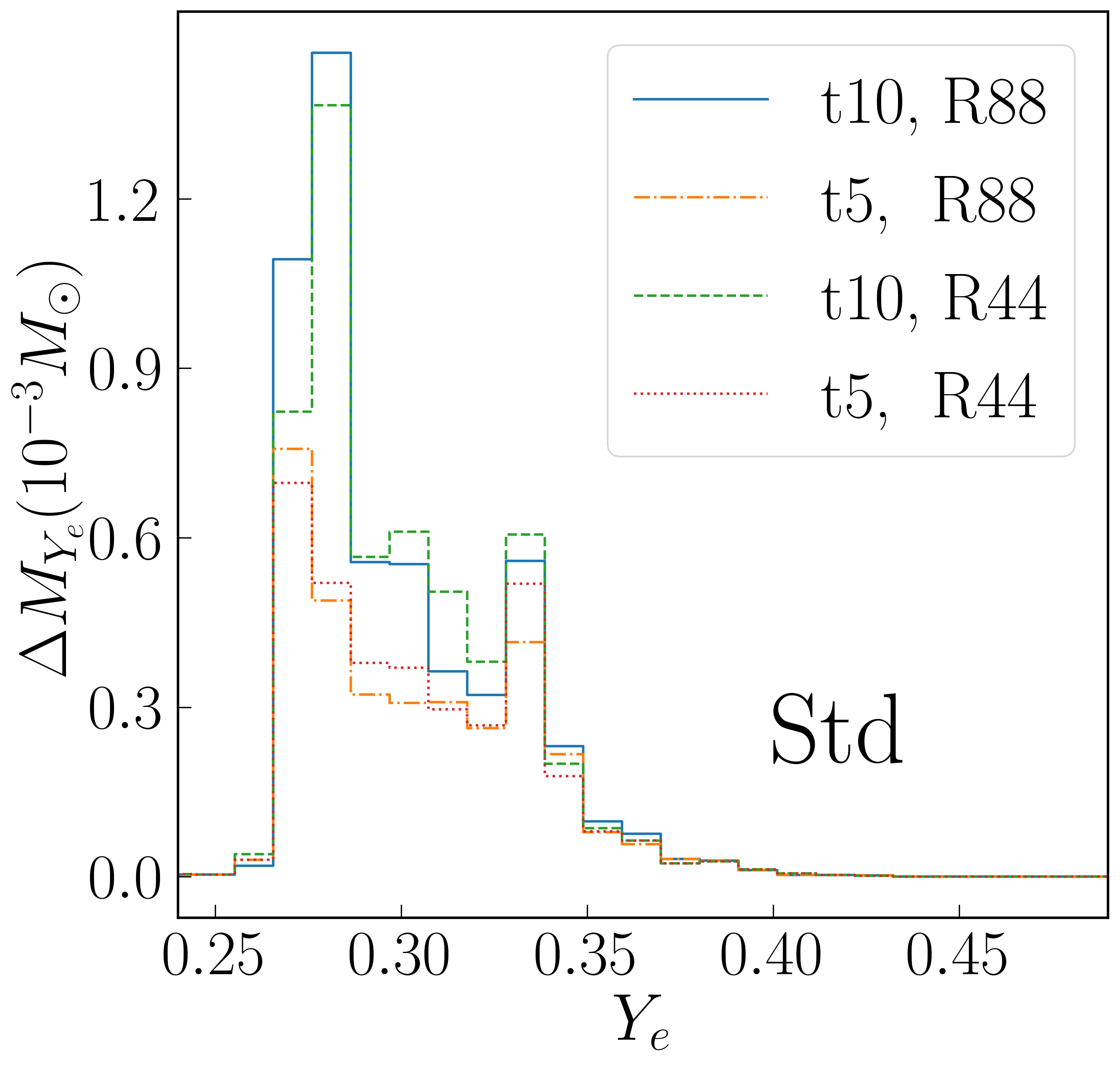}
  \includegraphics[width=8cm]{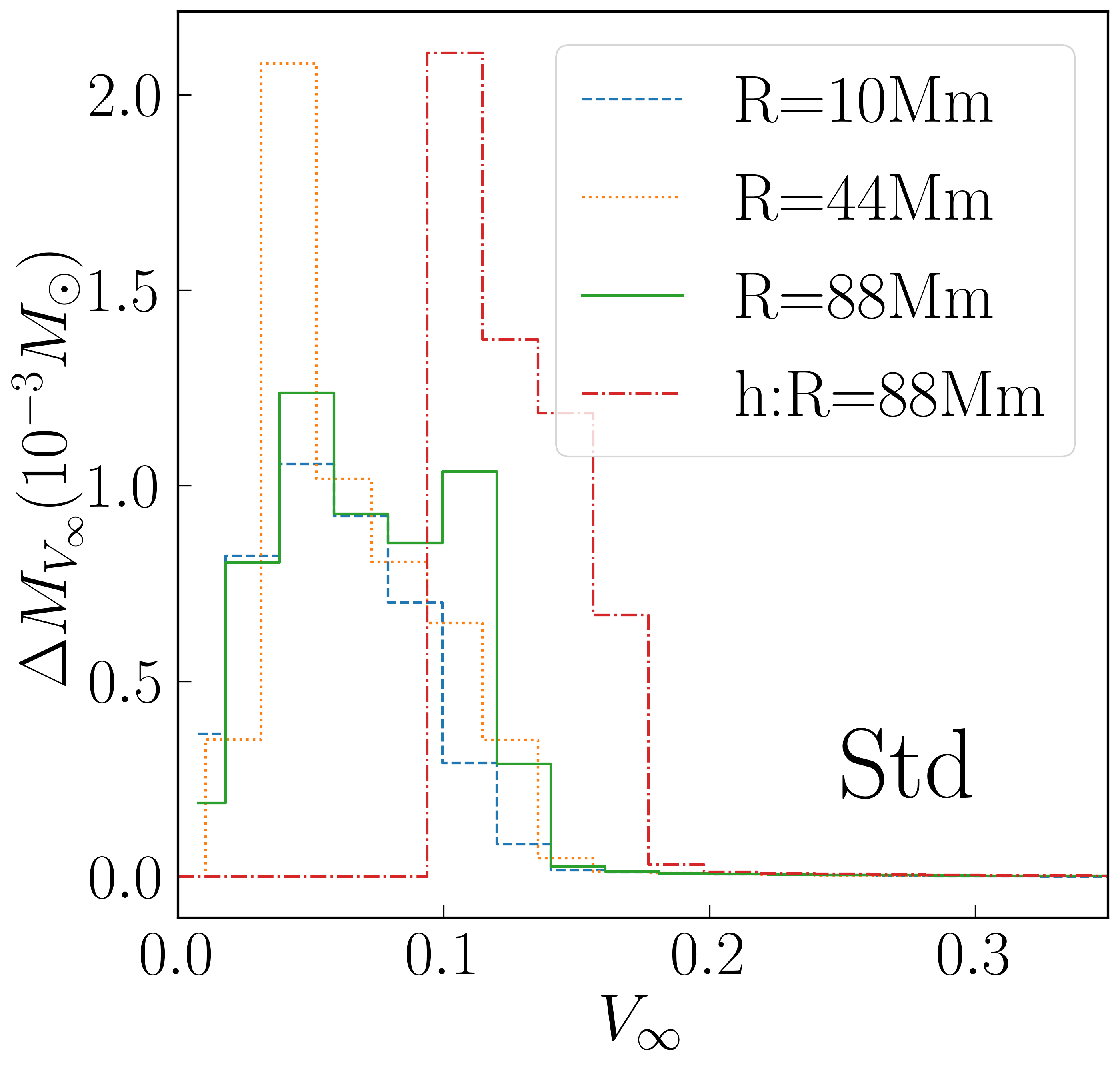} \\
  \includegraphics[width=8cm]{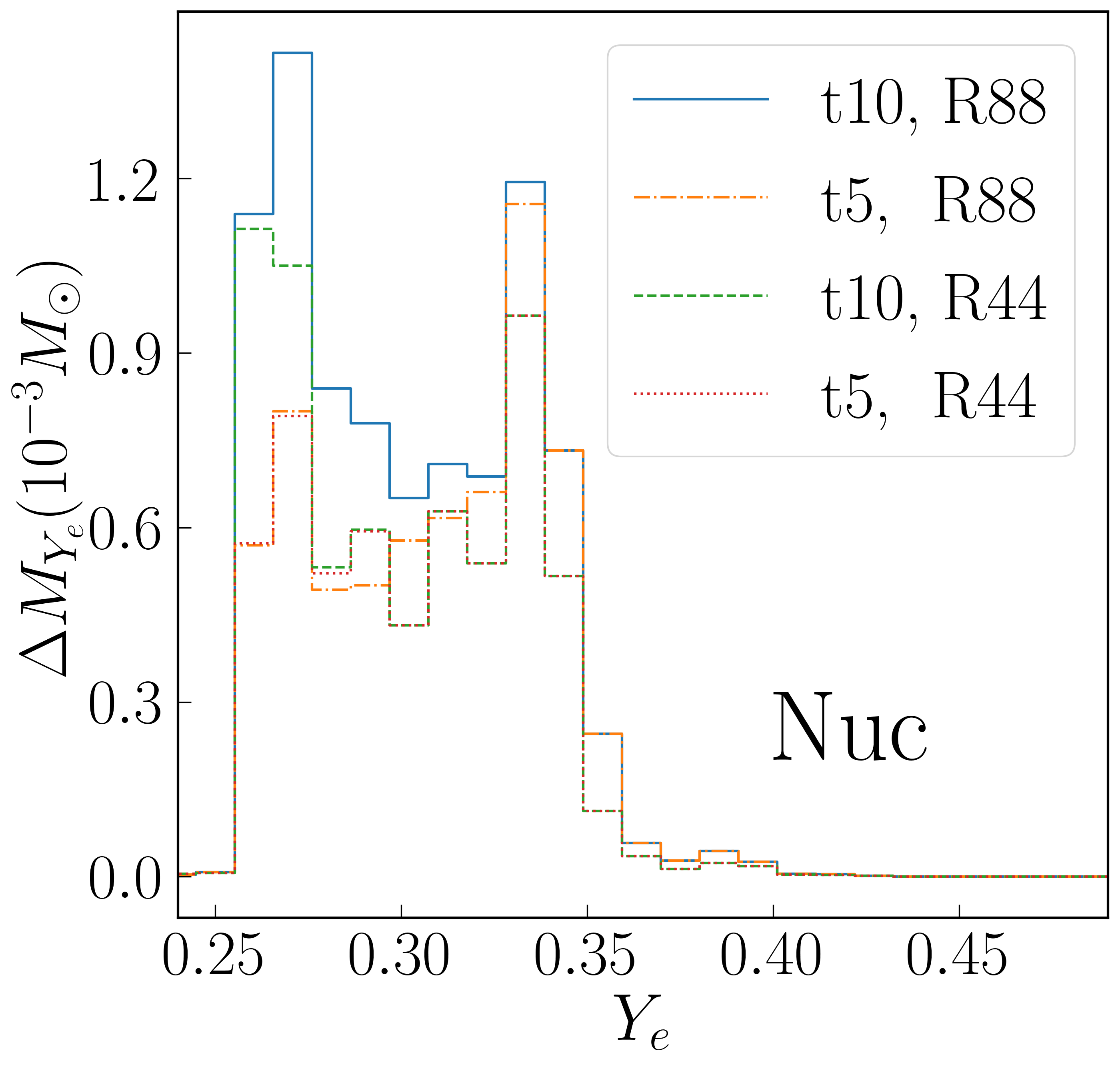}
  \includegraphics[width=8cm]{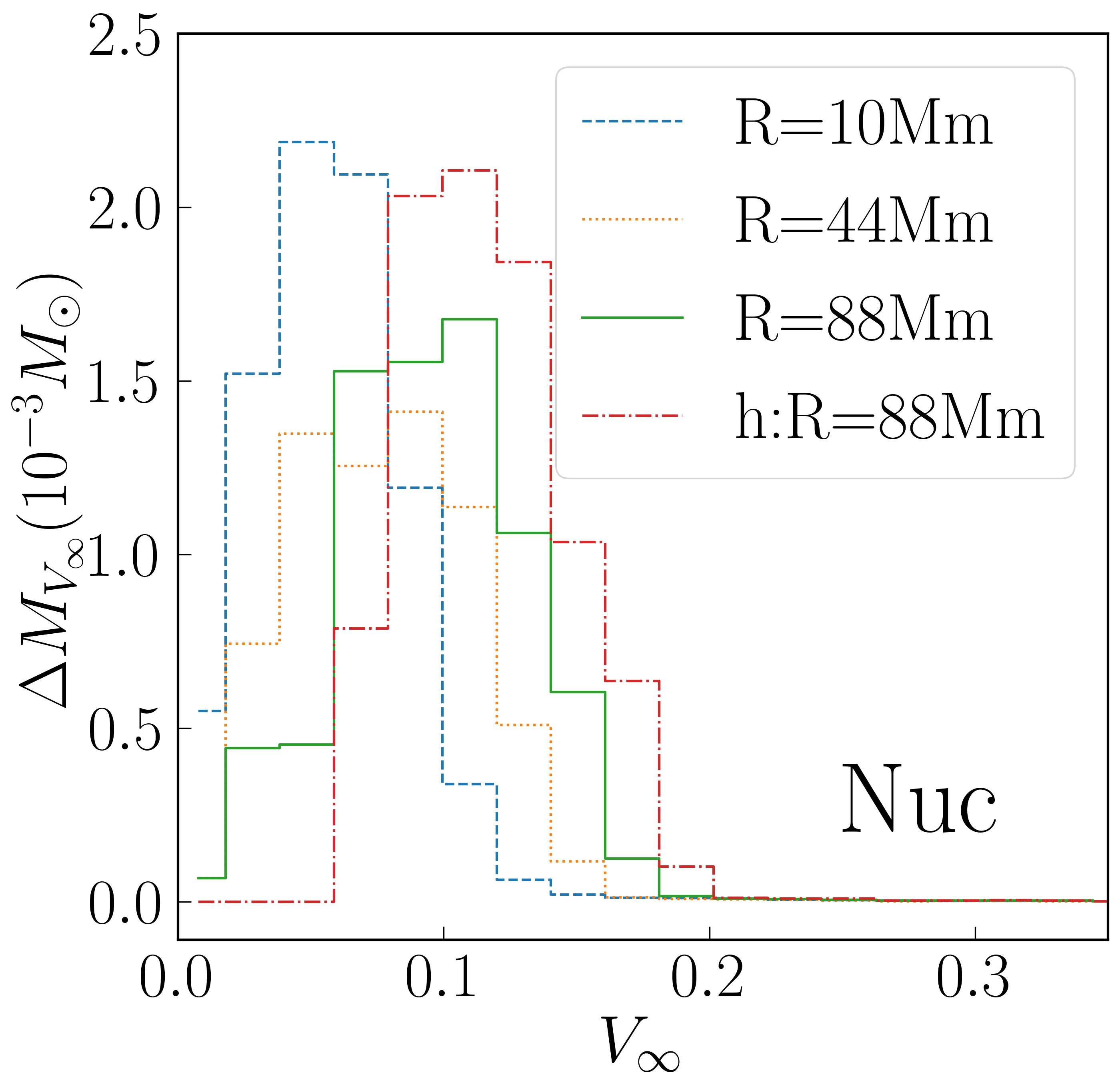}
  \caption
  {For the ``Std'' and ``Nuc'' runs, sensitivity of outflow distributions to the extraction time and radius.  The bin widths are $\Delta Y_e=0.01$, $\Delta V_{\infty}=0.02$.  The identity of the run in each panel is labeled in the lower right.
 }
  \label{fig:TandR}
\end{figure}

The $Y_e$ distributions are often found to exhibit a double peak structure.  In binary neutron star mergers, there is often a low-$Y_e$ equatorial ejecta and a higher-$Y_e$, more isotropic ejecta, but in our case all components of the ejecta with appreciable mass are equatorially concentrated.  Instead, the two peaks represent ejecta from different times.  As mentioned in the discussion of evolution of average $Y_e$ for bound matter, the ejecta from the first half of the evolution has a higher average $Y_e$, while the later ejecta has a narrower $Y_e$ distribution peaked at around 0.28.  An indication of this can be seen in the left-hand panel of Figure~\ref{fig:TandR}, which shows $Y_e$ ejecta distributions for Std at times of 5 and 10 seconds, and the effect of the time interval from 5 to 10 seconds is to accumulate outflow at the lower $Y_e$ peak (although even at 5 seconds, the late-time peak has begun to appear).

The r-process source terms have the anticipated effect of driving all the outflow in the rProc run to the target $Y_e$ i.e. the expected $Y_e$ at the end of the r-process). We note that this should not be understood as the rProc simulation predicting different values of $Y_e$ at the beginning of the r-process, and thus different r-process yields. Instead, the rProc simulation shows $Y_e$ at the end of the r-process, while all other simulations show $Y_e$ before r-process nucleosynthesis; the distributions of $Y_e$ at the end of these simulations are thus not directly comparable.

The simulated r-process does not increase the total mass of ejecta or its entropy, despite having deposited thermal energy into the gas.  In fact, these numbers are a bit lower for rProc than they are for Nuc. Just~{\it et al}~\cite{10.1093/mnras/stv009} also attempted to incorporate r-process effects by adding a heating term to outgoing fluid, and also found rather small effects on global outflow quantities.  That the effect in our case is to lower $M_{\rm out}$ and $\langle S\rangle$, however, remains counter-intuitive.  Possibly this is because the r-process timescale is similar to the timescale during which neutrinos are important, so during the first second, gas could undergo r-process alteration of $Y_e$ and radiated the resulting heat both from the $\dot{Q}$ of our model (Eq.~\ref{eq:Qdot}) and through the M1 neutrino transport active then, and this might greatly reduce the dominance of heating over cooling compared to what was anticipated from the analysis in Section~\ref{sec:rprocess}.  To check this, we re-ran Nuc from 1 to 5 seconds with r-process effects starting only at 1 second, after neutrino effects become negligible.  We find that adding the r-process source terms increases the outflow mass (defined as in Eq.~\ref{eq:Mout} but with the $t$ integral from 1 to 5) by 30\%.  This confirms that our r-process terms in a simpler setting do energize and unbind matter.  However, a reliable model of these effects will require a consistent treatment of neutrino losses.  Another expected effect of $r$-process heating on ejecta is a smoothing of spatial inhomogeneities, but this would not be visible on the length and timescales of this simulation~\cite{Rosswog:2013kqa,Fernandez:2014bra,Klion:2021jzr}.

In Figure~\ref{fig:TandR}, we investigate the sensitivity of our distributions to the observation radius and time.  $M_Y$ is insensitive to observation radius, since $Y_e$ of the outflow just advects at these late times.  However, we see that a significant fraction of the outflow is generated in the last half of the simulation, the last five seconds, so that evolution to at least ten seconds was indeed necessary.  The $V_{\infty}$ profiles are more sensitive to extraction radius.  This could already be seen in the higher $\langle V_{\infty}\rangle$ in Table~\ref{tab:ejecta}.  The question arises whether this acceleration is a physical effect that can be explained by pressure forces.  One way to estimate the possible importance of pressure forces is to compute the specific energy, and hence asymptotic velocity, from $h u_t/h(T=0)$ rather than from $u_t$.  Because the outflow is not steady, $h u_t$ is not a constant for fluid elements--it could not be, since for outflow of finite width, pressure forces must push fluid at the inner edge inward rather than outward--so this does not provide a more realistic asymptotic velocity, but it does indicate the magnitude of the effect pressure forces can have.  We see that pressure forces can quite easily push the asymptotic velocity above 0.1\,$c$.  Another concern is whether pressure could be anomalously high in our simulations.  However, the average specific entropy reported in Table~\ref{tab:ejecta} is well in line with entropy reported in other studies~\cite{Fernandez:2020oow,Fujibayashi:2020jfr}.
Thus, it is likely that the observation radius of $r_{\rm ob}=$10,000\,km, which is fairly common in prior studies, underestimates the asymptotic velocities in earlier works, even apart from nuclear heating effects which were known to be missing.  Confirmation of this interpretation comes from the study of Kasen~{\it et al}~\cite{Kasen:2014toa}, in which disk outflow at $10^4$\,km was injected onto a much larger grid, and it was found that homology was not achieved until $\sim 100$\,s.

Strong acceleration from $10^4$\,km and $10^5$\,km is present in Nuc as well as in Std, as clearly seen from Table~\ref{tab:ejecta} and Figure~\ref{fig:TandR}.  For Nuc, there seems to be less internal energy available for acceleration beyond $r>10^5$\,km than there is for Std.  The asymptotic velocity using $h u_t$ is similar in both cases, but the outflow does more of its acceleration inside $10^5$\,km in the Nuc case.

\section{Discussion}
\label{sec:discussion}

By simulating the post-merger evolution of a black hole-neutron star
binary, we have investigated the importance of different physical
effects on the late-time evolution of the disk and the properties of
outflows.  At least for $\alpha_{\rm visc}=0.03$, evolution for nearly ten seconds is needed for a full accounting of the outflow.  Evolution of the remnant beyond this point is unnecessary, since less than $10^{-3}M_{\odot}$ of bound material remains.

Large spatial extents are needed to track the expanding disk and outflow; our computational grid extended to about $10^5$\,km, which was adequate to contain the disk but did not observe outflow trajectories become ballistic.  This is probably our most important conclusion, that asymptotic velocities of ejecta in simulations with boundaries like ours (i.e. most simulations) are not reliable, that the asymptotic velocity is likely to be significantly greater than the $\sim 0.05c$ often reported.  Indications of this have appeared in earlier work~\cite{Kasen:2014toa}, but it has not been emphasized and appreciation of it is still spreading through the relevant astrophysics community.

Our conclusion about the large distance to the homologous regime ironically suggests we move our outer boundaries inward in future simulations, gaining radial resolution at the cost of grid extent.  Capturing the entire range from the black hole to the homology regime on a single radially logarithmic grid is impractical, since the timestep is restricted by the high resolution near the black hole.  One should set boundaries far enough that only unbound material reaches it during the first ten seconds, monitor outflow, and use this as the inflow inner boundary condition for a larger grid simulation, as was done by Kasen~{\it et al}~\cite{Kasen:2014toa}.

A second important conclusion is that the inclusion of heavy nuclei is important for the outflow properties; in fact, it made the single largest difference in ejecta mass and entropy of any variation.   Given the 60\% difference in outflow mass, it will be worthwhile for future simulations to incorporate a more realistic nuclei mix than the free nucleons and alpha particles model.  The effect of r-process heating was noticeably smaller by comparison, but this statement must be heavily qualified by acknowledging the simplicity and imperfections of our approximate treatment of the r-process.  The fact that an ejecta energy boost was not seen in the run with r-process effects may indicate the care needed in dealing with neutrino cooling feedback in models with approximate r-process modeling.

A third major conclusion is on the likely nature of composition and entropy transport effects, which are automatically accounted in turbulent MHD simulations but usually omitted in viscous disk models.  The main effect seems to be to drain energy from the higher-energy material that would become unbound, so that outflows have lower mass and velocity than would otherwise occur.  Our treatment of these transport effects is probably only qualitatively correct, but it is likely that the signs of their influences on outflows actually is what we find.  The $Y_e$ distribution of the outflow, by contrast, was only mildly changed.  This probably indicates that heat transport, which has the bigger influence on energetics, is much more important than particle diffusion, which can more safely be ignored.

The dependencies on nuclear and neutrino physics suggest that further improvements to their treatment should be pursued.  The grey M1 neutrino transport scheme used for this study, while an improvement over neutrino leakage, remains very imperfect.  Aside from the average energy per neutrino (retained by separately evolving neutrino energy density and number density), information on the neutino energy spectra is lost.  Most angle dependence in the distribution functions is also discarded.  The fixed closure condition does not allow convergence to the true solution to the transport equation, and it is known to produce artificial ``radiation shocks'' on the poles above radiating disks, which we do see in our simulations.  It thus cannot reliably estimate neutrino-antineutrino annihilation rates in this region.  Fortunately, SpEC's newer full neutrino radiation transport method using Monte Carlo techniques overcomes these problems~\cite{Foucart:2021mcb}.  We are currently working to apply these methods to late-time axisymmetric simulations.  Only with results from simulations evolving Boltzmann's equation of radiation transport.

It has long been known that alpha particle recombination is important for understanding post-merger disk winds( e.g.~\cite{2009ApJ...699L..93L}), and we have seen that inclusion of heavy nuclei and r-process protonization also have sufficiently large effect for their inclusion to be worthwhile.  Because of the low densities and temperatures involved, a fully reliable treatment would have to dispense with the wonderful simplification of nuclear statistical equilibrium.  The nuclear reaction networks used to track nuclear reactions in tracer particles from post-merger simulations evolve many thousands of isotope abundances, which would be impractical for even two-dimensional simulations.  Whether evolving a much smaller set of abundances of isotope groups would be sufficient for energetic purposes (as opposed to detailed nucleosynthetic abundance output) deserves consideration.

Given the challenges of capturing the physics of post-merger evolution, axisymmetric simulations will surely remain useful for some time to come.  This may involve modeling transport effects using viscosity and diffusion, by dynamo-enhanced magnetohydrodynamics, or by some combination of the two, depending on what detailed calibration to 3D MHD simulations and developments in the theory of accretion disk dynamos reveals.

\ack
We thank Matthias Hempel for assistance in the use of his equation of state tables.  M.D. gratefully acknowledges support from the NSF through grant PHY-2110287.  M.D. and F.F. acknowledge support from NASA through grant 80NSSC22K0719. 
F.F. and A.K. acknowledge support from the DOE through grant DE-SC0020435,
 and from NASA through grant 80NSSC18K0565.
R.F acknowledges support from the Natural Sciences and Engineering
Research Council (NSERC) of Canada through grant
RGPIN-2022-03463.
J.J. acknowledges support from the Washington NASA
Space Grant Consortium, NASA Grant NNX15AJ98H.
L.K. acknowledges support form the Sherman Fairchild Foundation and NSF Grants No. PHY-1912081 and No. OAC-1931280 at Cornell.
M.S. acknowledges support from the Sherman Fairchild Foundation and NSF Grants No. PHY-2011961, PHY-2011968, and NSF-OAC-1931266 at Caltech.
\newline

\bibliographystyle{iopart-num}
\bibliography{late-time-ref}

\providecommand{\newblock}{}
\begin{thebibliography}{10}
\expandafter\ifx\csname url\endcsname\relax
  \def\url#1{{\tt #1}}\fi
\expandafter\ifx\csname urlprefix\endcsname\relax\def\urlprefix{URL }\fi
\providecommand{\eprint}[2][]{\url{#2}}

\bibitem{LIGOScientific:2017vwq}
Abbott B~P {\em et~al.\/} (LIGO Scientific, Virgo) 2017 {\em Phys. Rev.
  Lett.\/} {\bf 119} 161101 (\textit{Preprint} \eprint{1710.05832})

\bibitem{LIGOScientific:2020aai}
Abbott B~P {\em et~al.\/} (LIGO Scientific, Virgo) 2020 {\em Astrophys. J.
  Lett.\/} {\bf 892} L3 (\textit{Preprint} \eprint{2001.01761})

\bibitem{LIGOScientific:2021qlt}
Abbott R {\em et~al.\/} (LIGO Scientific, KAGRA, VIRGO) 2021 {\em Astrophys. J.
  Lett.\/} {\bf 915} L5 (\textit{Preprint} \eprint{2106.15163})

\bibitem{Lippuner2015}
{Lippuner} J and {Roberts} L~F 2015 {\em Astrophys. J.\/} {\bf 815} 82
  (\textit{Preprint} \eprint{1508.03133})

\bibitem{Metzger:2019zeh}
Metzger B~D 2020 {\em Living Rev. Rel.\/} {\bf 23} 1 (\textit{Preprint}
  \eprint{1910.01617})

\bibitem{2018ApJ...858...52S}
{Siegel} D~M and {Metzger} B~D 2018 {\em Astrophys.\ J.\/} {\bf 858} 52
  (\textit{Preprint} \eprint{1711.00868})

\bibitem{2017PhRvL.119w1102S}
{Siegel} D~M and {Metzger} B~D 2017 {\em Phys.\ Rev.\ Lett.\/} {\bf 119} 231102
  (\textit{Preprint} \eprint{1705.05473})

\bibitem{10.1093/mnras/sty2932}
Fernández R, Tchekhovskoy A, Quataert E, Foucart F and Kasen D 2018 {\em
  Monthly Notices of the Royal Astronomical Society\/} {\bf 482} 3373--3393
  ISSN 0035-8711 (\textit{Preprint}
  \eprint{https://academic.oup.com/mnras/article-pdf/482/3/3373/26660289/sty2932.pdf})
  \urlprefix\url{https://doi.org/10.1093/mnras/sty2932}

\bibitem{PhysRevD.100.023008}
Miller J~M, Ryan B~R, Dolence J~C, Burrows A, Fontes C~J, Fryer C~L, Korobkin
  O, Lippuner J, Mumpower M~R and Wollaeger R~T 2019 {\em Phys. Rev. D\/} {\bf
  100}(2) 023008
  \urlprefix\url{https://link.aps.org/doi/10.1103/PhysRevD.100.023008}

\bibitem{2021arXiv211104621H}
{Hayashi} K, {Fujibayashi} S, {Kiuchi} K, {Kyutoku} K, {Sekiguchi} Y and
  {Shibata} M 2021 {\em arXiv e-prints\/} arXiv:2111.04621 (\textit{Preprint}
  \eprint{2111.04621})

\bibitem{1933MNRAS..94...39C}
{Cowling} T~G 1933 {\em Mon.\ Not.\ Roy.\ Astr.\ Soc.\/} {\bf 94} 39--48

\bibitem{Shibata:2021xmo}
Shibata M, Fujibayashi S and Sekiguchi Y 2021 {\em Phys. Rev. D\/} {\bf 104}
  063026 (\textit{Preprint} \eprint{2109.08732})

\bibitem{shakura:1973}
{Shakura} N~I and {Sunyaev} R~A 1973 {Black Holes in Binary Systems:
  Observational Appearances} {\em X- and Gamma-Ray Astronomy\/} ({\em IAU
  Symposium\/} vol~55) ed {Bradt} H and {Giacconi} R p 155

\bibitem{Fernandez2013}
{Fern{\'a}ndez} R and {Metzger} B~D 2013 {\em Mon.\ Not.\ Roy.\ Astr.\ Soc.\/}
  {\bf 435} 502--517 (\textit{Preprint} \eprint{1304.6720})

\bibitem{Fernandez:2014}
{Fern{\'a}ndez} R, {Kasen} D, {Metzger} B~D and {Quataert} E 2015 {\em Mon.\
  Not.\ Roy.\ Astr.\ Soc.\/} {\bf 446} 750--758 (\textit{Preprint}
  \eprint{1409.4426})

\bibitem{Fernandez:2014b}
Fern{\'a}ndez R, Quataert E, Schwab J, Kasen D and Rosswog S 2015 {\em Mon.
  Not. Roy. Astron. Soc.\/} {\bf 449} 390--402 (\textit{Preprint}
  \eprint{1412.5588})

\bibitem{Fernandez:2016sbf}
Fern{\'a}ndez R, Foucart F, Kasen D, Lippuner J, Desai D and Roberts L~F 2017
  {\em Class. Quant. Grav.\/} {\bf 34} 154001 (\textit{Preprint}
  \eprint{1612.04829})

\bibitem{Fernandez:2020oow}
Fern\'andez R, Foucart F and Lippuner J 2020 {\em Mon. Not. Roy. Astron.
  Soc.\/} {\bf 497} 3221--3233 (\textit{Preprint} \eprint{2005.14208})

\bibitem{10.1093/mnras/stv009}
Just O, Bauswein A, Pulpillo R~A, Goriely S and Janka H~T 2015 {\em Monthly
  Notices of the Royal Astronomical Society\/} {\bf 448} 541--567 ISSN
  0035-8711 (\textit{Preprint}
  \eprint{https://academic.oup.com/mnras/article-pdf/448/1/541/9387243/stv009.pdf})
  \urlprefix\url{https://doi.org/10.1093/mnras/stv009}

\bibitem{PhysRevD.101.083029}
Fujibayashi S, Shibata M, Wanajo S, Kiuchi K, Kyutoku K and Sekiguchi Y 2020
  {\em Phys. Rev. D\/} {\bf 101}(8) 083029
  \urlprefix\url{https://link.aps.org/doi/10.1103/PhysRevD.101.083029}

\bibitem{Fujibayashi:2020jfr}
Fujibayashi S, Shibata M, Wanajo S, Kiuchi K, Kyutoku K and Sekiguchi Y 2020
  {\em Phys. Rev. D\/} {\bf 102} 123014 (\textit{Preprint} \eprint{2009.03895})

\bibitem{Fernandez:2014bra}
Fern\'andez R, Quataert E, Schwab J, Kasen D and Rosswog S 2015 {\em Mon. Not.
  Roy. Astron. Soc.\/} {\bf 449} 390--402 (\textit{Preprint}
  \eprint{1412.5588})

\bibitem{Fujibayashi:2017puw}
Fujibayashi S, Kiuchi K, Nishimura N, Sekiguchi Y and Shibata M 2018 {\em
  Astrophys. J.\/} {\bf 860} 64 (\textit{Preprint} \eprint{1711.02093})

\bibitem{Armengol:2021mbt}
Armengol F~G~L {\em et~al.\/} 2021  (\textit{Preprint} \eprint{2112.09817})

\bibitem{10.1111/j.1365-2966.2009.16107.x}
Metzger B~D, Arcones A, Quataert E and Martínez-Pinedo G 2010 {\em Monthly
  Notices of the Royal Astronomical Society\/} {\bf 402} 2771--2777 ISSN
  0035-8711 (\textit{Preprint}
  \eprint{https://academic.oup.com/mnras/article-pdf/402/4/2771/4912275/mnras0402-2771.pdf})
  \urlprefix\url{https://doi.org/10.1111/j.1365-2966.2009.16107.x}

\bibitem{Duez:2020lgq}
Duez M~D, Knight A, Foucart F, Haddadi M, Jesse J, Hebert F, Kidder L~E,
  Pfeiffer H~P and Scheel M~A 2020 {\em Phys. Rev. D\/} {\bf 102} 104050
  (\textit{Preprint} \eprint{2008.05019})

\bibitem{Foucart:2014nda}
{Foucart} F, {Deaton} M~B, {Duez} M~D, {O'Connor} E, {Ott} C~D, {Haas} R,
  {Kidder} L~E, {Pfeiffer} H~P, {Scheel} M~A and {Szil{\'a}gyi} B 2014 {\em
  Phys.\ Rev.\ D\/} {\bf 90} 024026 (\textit{Preprint} \eprint{1405.1121})

\bibitem{Lattimer:1991nc}
Lattimer J~M and Swesty F~D 1991 {\em Nucl. Phys.\/} {\bf A535} 331--376

\bibitem{SpEC2020}
{SpEC}: Spectral einstein code \url{https://www.black-holes.org/code/SpEC.html}
  [{A}ccessed Feb. 25, 2020]

\bibitem{Jesse:2020}
Jesse J, Duez M~D, Foucart F, Haddadi M, Knight A~L, Cadenhead C~L,
  H{\'{e}}bert F, Kidder L~E, Pfeiffer H~P and Scheel M~A 2020 {\em Classical
  and Quantum Gravity\/} {\bf 37} 235010
  \urlprefix\url{https://doi.org/10.1088%2F1361-6382%2Fabbc8b}

\bibitem{Nouri:2017fvh}
{Nouri} F~H, {Duez} M~D, {Foucart} F, {Deaton} M~B, {Haas} R, {Haddadi} M,
  {Kidder} L~E, {Ott} C~D, {Pfeiffer} H~P, {Scheel} M~A and {Szilagyi} B 2018
  {\em Phys. Rev. D\/} {\bf 97} 083014 (\textit{Preprint} \eprint{1710.07423})

\bibitem{Radice:2017}
{Radice} D 2017 {\em Astrophys.\ J.\ Lett.\/} {\bf 838} L2 (\textit{Preprint}
  \eprint{1703.02046})

\bibitem{1999MNRAS.303..309I}
{Igumenshchev} I~V and {Abramowicz} M~A 1999 {\em Mon.\ Not.\ Roy.\ Astr.\
  Soc.\/} {\bf 303} 309--320

\bibitem{FoucartM1:2015}
{Foucart} F, {O'Connor} E, {Roberts} L, {Duez} M~D, {Haas} R, {Kidder} L~E,
  {Ott} C~D, {Pfeiffer} H~P, {Scheel} M~A and {Szil{\'a}gyi} B 2015 {\em Phys.\
  Rev.\ D\/} {\bf 91} 124021 (\textit{Preprint} \eprint{1502.04146})

\bibitem{Foucart:2016rxm}
{Foucart} F, {O'Connor} E, {Roberts} L, {Kidder} L~E, {Pfeiffer} H~P and
  {Scheel} M~A 2016 {\em Phys. Rev.\/} {\bf D94} 123016 (\textit{Preprint}
  \eprint{1607.07450})

\bibitem{Timmes:2000}
{Timmes} F~X and {Swesty} F~D 2000 {\em Astrophys.\ J. Suppl. Ser.\/} {\bf 126}
  501--516

\bibitem{1996ApJS..106..171B}
{Blinnikov} S~I, {Dunina-Barkovskaya} N~V and {Nadyozhin} D~K 1996 {\em
  Astrophys.\ J. Suppl. Ser.\/} {\bf 106} 171

\bibitem{Hempel:2011mk}
Hempel M, Fischer T, Schaffner-Bielich J and Liebend{\"o}rfer M 2012 {\em
  Astrophys.J.\/} {\bf 748} 70 (\textit{Preprint} \eprint{1108.0848})

\bibitem{hempel_2015}
Hempel M 2015 Equations of state
  \urlprefix\url{https://astro.physik.unibas.ch/en/people/matthias-hempel/equations-of-state/}

\bibitem{Foucart:2021ikp}
Foucart F, Moesta P, Ramirez T, Wright A~J, Darbha S and Kasen D 2021 {\em
  Phys. Rev. D\/} {\bf 104} 123010 (\textit{Preprint} \eprint{2109.00565})

\bibitem{Wu:2016pnw}
Wu M~R, Fern\'andez R, Mart\'\i{}nez-Pinedo G and Metzger B~D 2016 {\em Mon.
  Not. Roy. Astron. Soc.\/} {\bf 463} 2323--2334 (\textit{Preprint}
  \eprint{1607.05290})

\bibitem{AccretionPower}
{Frank} J, {King} A and {Raine} D~J 2002 {\em {Accretion Power in Astrophysics:
  Third Edition}\/}

\bibitem{Beloborodov:2002af}
Beloborodov A~M 2003 {\em Astrophys. J.\/} {\bf 588} 931--944
  (\textit{Preprint} \eprint{astro-ph/0210522})

\bibitem{Siegel2017}
{Siegel} D~M and {Metzger} B~D 2017 {\em ArXiv e-prints\/} (\textit{Preprint}
  \eprint{1705.05473})

\bibitem{Rosswog:2013kqa}
Rosswog S, Korobkin O, Arcones A, Thielemann F~K and Piran T 2014 {\em Mon.
  Not. Roy. Astron. Soc.\/} {\bf 439} 744--756 (\textit{Preprint}
  \eprint{1307.2939})

\bibitem{Klion:2021jzr}
Klion H, Tchekhovskoy A, Kasen D, Kathirgamaraju A, Quataert E and Fern\'andez
  R 2022 {\em Mon. Not. Roy. Astron. Soc.\/} {\bf 510} 2968--2979
  (\textit{Preprint} \eprint{2108.04251})

\bibitem{Kasen:2014toa}
Kasen D, Fernandez R and Metzger B 2015 {\em Mon. Not. Roy. Astron. Soc.\/}
  {\bf 450} 1777--1786 (\textit{Preprint} \eprint{1411.3726})

\bibitem{Foucart:2021mcb}
Foucart F, Duez M~D, Hebert F, Kidder L~E, Kovarik P, Pfeiffer H~P and Scheel
  M~A 2021 {\em Astrophys. J.\/} {\bf 920} 82 (\textit{Preprint}
  \eprint{2103.16588})

\bibitem{2009ApJ...699L..93L}
{Lee} W~H, {Ramirez-Ruiz} E and {L{\'o}pez-C{\'a}mara} D 2009 {\em Astrophys.\
  J.\ Lett.\/} {\bf 699} L93--L96 (\textit{Preprint} \eprint{0904.3752})

\end{thebibliography}

\end{document}